\documentclass[preprint,showpacs,preprintnumbers,amsmath,amssymb]{revtex4}
\usepackage{amsfonts}
\usepackage{amssymb}
\usepackage{latexsym}
\usepackage{fancybox}
\usepackage{graphicx}
\usepackage{psfig}
\newcommand{\al}{\alpha}
\newcommand{\be}{\beta}

\newcommand{\si}{\sigma}

\newcommand{\ga}{\gamma}

\newcommand{\de}{\delta}

\newcommand{\De}{\Delta}

\newcommand{\tha}{\theta}

\newcommand{\rar}{\rightarrow}

\begin{document}

\preprint{M\'exico ICN-UNAM 15/04 (2004) \
                           October 2006 (revised)}

\title{Exotic molecular ions $(HeH)^{2+}$ and $He_2^{3+}$ in a strong
magnetic field: low-lying states}

\author{A.V.~Turbiner}

\email{turbiner@nuclecu.unam.mx}
\author{J.C. \surname{L\'opez Vieyra}}
\email{vieyra@nuclecu.unam.mx}
\affiliation{}
\affiliation{Instituto de Ciencias Nucleares, UNAM, Apartado
Postal 70-543, 04510 M\'exico}


\begin{abstract}
  The Coulombic systems $(\al p e)$ and $(\al\al e)$, $(\al p p e)$,
  $(\al \al p e)$ and $(Li^{3+} Li^{3+} e)$ placed in a magnetic field
  $B \gtrsim 10^{11}\,\mbox{G}$ are studied.  It is demonstrated a
  theoretical existence of the exotic ion $(He H)^{2+}$ for
  $B\gtrsim 5\times 10^{12}\,\mbox{G}$ in parallel configuration
  (the magnetic field is directed along internuclear axis) as optimal
  as well as its excited states $1\pi, 1\delta$. As for the exotic ion
  ${He}_{2}^{3+}$ it is shown that in spite of strong
  electrostatic repulsion of $\al-$particles this ion can also exist for $B
  \gtrsim 100\,\mbox{a.u.} (= 2.35\times 10^{11}\,\mbox{G})$ in
  parallel configuration as optimal in the states $1\si_g$ (ground
  state), $1\pi_u, 1\delta_g$.  Upon appearance both ions are unstable
  towards dissociation with $He^+$ in the final state but with very
  large lifetime. However, at $B\gtrsim 10000$\,a.u. the ion $(He H)^{2+}$
  becomes stable, while at $B\gtrsim 1000$\,a.u. the ion ${He}_{2}^{3+}$
  becomes stable. With a magnetic field growth, both exotic ions become
  more and more tightly bound and compact, their lowest rotational and
  vibrational energies grow. At the edge of applicability of
  non-relativistic approximation, $B \sim 4.414 \times 10^{13}$\,G,
  there are indications that three more exotic linear ions $(H-He-H)^{3+}$,
  $(He-H-He)^{4+}$ and even $Li_2^{5+}$ in parallel configuration may also occur.
\end{abstract}

\pacs{31.15.Pf,31.10.+z,32.60.+i,97.10.Ld}

\maketitle


The possibility of formation of new exotic chemical compounds which
do not exist in field-free case is one of the most fascinating
features of the physics in strong magnetic fields \cite
{Kadomtsev:1971,Ruderman:1971}. In particular, it was conjectured
long ago that linear neutral chains of molecules (polymer-type
molecules) situated along a magnetic line might exist in
astro-physically strong magnetic fields. They were called the {\it
Ruderman chains}. In the year 1999, the first theoretical indication
for the existence of the exotic one-electron molecular ion
$H_3^{2+}$ in magnetic fields $B\gtrsim 10^{11}$\,G was provided
\cite{Turbiner:1999}. It is a sufficiently long-living state for
which the optimal configuration is parallel (all protons situated
along a magnetic line) decaying to $H_2^{+}+p$ \footnote{Detailed
study is presented in \cite{Turbiner:2004b}}. Perhaps, the most
astonishing result is an observation of the fast increase of the
ionization energy of the exotic $H_3^{2+}$ ion with a magnetic field
growth. Eventually, for magnetic fields $B\gtrsim 3 \times
10^{13}$\,G the binding energy of the exotic $H_3^{2+}$ ion becomes
larger than the binding energy of the traditional $H_2^{+}$ ion. As
a result the exotic ion $H_3^{2+}$ becomes stable as well as the
traditional ion $H_2^{+}$ and the hydrogen atom. Later in the year
2000, it was indicated the possible existence of the exotic
one-electron molecular ion $H_4^{3+}$ for magnetic fields $B\gtrsim
10^{13}$\,G \cite{Lopez-Tur:2000}. It is worth emphasizing that for
{\it all} studied one-electron molecular systems $H_2^{+}, H_3^{2+},
H_4^{3+}$ the optimal configuration for $B\gtrsim 10^{11}$\,G is
parallel, when all protons were situated along a magnetic line (see,
in particular, \cite{Turbiner:2002}). Very recently, it was
announced that two $\al$-particle contained ions, $(He H)^{2+}$ and
${He}_{2}^{3+}$ can exist at $B\gtrsim 10^{12}$ \cite{Turbiner:2005}
(for a discussion and review the status of one-electron systems, see
\cite{Turbiner:2006}).

Usually, the investigations of the Coulomb systems in a strong
magnetic field were justified by the fact there exists a strong
magnetic field on a surface of neutron stars or some white dwarfs.
The recent discovery of absorption features at 0.7\,KeV and 1.4\,KeV
in the X-ray spectrum of the isolated neutron star 1E1207.4-5209 by
Chandra $X$-ray observatory \cite{Sanwal:2002} and its further
confirmation by XMM-Newton $X$-ray observatory \cite{Bignami:2003}
(see also \cite{Mori:2005}) motivated more profound and extended
studies in a strong magnetic field. In particular, it was assumed
the hydrogenic atmosphere of 1E1207.4-5209 with main abundance of
the exotic ion $H_3^{2+}$ \cite{Turbiner:2004m} which may explain
the above absorption features. There were already found two other
neutron stars whose atmospheres are characterized by absorption
features \cite{Kerkwijk:2004,Vink:2004}, which are waiting to be
explained. These discoveries pushed us to make a detailed study of
traditional atomic-molecular systems as well as for a search for new
exotic chemical compounds in a strong magnetic field.

The goal of present article is to make a detailed investigation of
Coulomb systems made from one electron and several protons and/or
$\al$-particles in presence of a strong magnetic field. It is
natural to expect that due to vanishing rotational energy the
optimal configuration with minimal total energy is the parallel
configuration: all massive positively-charged particles are situated
along a magnetic line. We demonstrate that for sufficiently strong
magnetic fields which appear on the surfaces of the neutron stars
the exotic ions $(HeH)^{2+}$, $He_2^{3+}$ can exist, while the ions
$(H-He-H)^{3+}$, $(He-H-He)^{4+}$ and $Li_2^{5+}$ may appear to
exist as well. The consideration is non-relativistic, based on a
variational solution of the Schroedinger equation. Hence, the
magnetic field strength is restricted by the Schwinger limit
$B=4.414\times 10^{13}$\,G. Also it is based on the Born-Oppenheimer
approximation of zero order: the particles of positive charge
(protons and $\al$-particles) are assumed to be infinitely massive.

Atomic units are used throughout ($\hbar$=$m_e$=$e$=1), albeit
energies are expressed in Rydbergs (Ry). The magnetic field $B$ is
given in a.u. with $B_0= 2.35 \times 10^9$\,G.

\section{$(He H)^{2+}$ molecular ion}

Since long ago the hybrid system $(He H)^{2+}$, made out of
$\al$-particle, proton and electron, $(\al p e)$, was attempted to
explore for field-free case \cite{Winter:1977} and for the case of a
magnetic field of moderate strength $B=1$\,a.u. \cite{Wille:1988}.
In both studies no indication to appearance of a bound state was
observed. Here we will show that for larger magnetic fields
$B\gtrsim 10^{12}$\,G the exotic hybrid ion $(He H)^{2+}$ can exist
in parallel configuration as optimal being a sufficiently
long-living state. It decays to $He^{+} + p$.  For all magnetic
fields the binding energy of $(He H)^{2+}$ is slightly smaller than
the binding energy of the atomic ion $He^{+}$ and their difference
decreases as a magnetic field grows. In parallel configuration the
orbital momentum projection on the molecular axis is preserved and
eigenstates are characterized by magnetic quantum number $m$, for
the ground state $m=0$ (Perron theorem).

The Hamiltonian which describes two infinitely heavy centers $a$ and
$b$ of charges $Z_1$ and $Z_2$ situated along the line forming the
angle $\tha$ with the $z-$axis, and electron placed in a uniform
constant magnetic field directed along the $z-$axis, $\nobreak{{\bf
B}=(0,0,B)}$ is given by
\begin{equation}
 \label{ham-HeH}
 {\cal H} = -\De  -
\frac{2\,Z_1}{r_1} -\frac{2 Z_2}{r_2}\,+\frac{2\,Z_1 Z_2}{R}
  + ({\hat p} {\cal A}+{\cal A}{\hat p}) +  {\cal A}^2 \ ,
\end{equation}
(for geometrical setting see Fig.~1). In the case of $(He H)^{2+}$
the charges $Z_1=Z=2$ and $Z_2=1$ correspond to the
$\alpha-$particle and the proton as heavy charged centers. The
vector potential is given by a certain one-parameter family of
vector potentials corresponding to a constant magnetic field ${\bf
B}=(0,0,B)$
\begin{equation}
\label{Vec}
  {\cal A}= B((\xi-1)y,\ \xi x,\ 0)\ ,
\end{equation}
where $\xi$ is a parameter. The position of the {\it gauge center}
or {\it gauge origin}, where $\nobreak{{\cal A}(x,y,z)=0}$, is
defined by $x=y=0$, with $z$ arbitrary. For simplicity we fix $z=0$.
The gauge origin $O$ is chosen to be located somewhere along the
line connecting the charged centers but not necessarily coinciding
with the mid-point $O'$ (see Fig.~\ref{fig:1}).  If $\xi=1/2$ we get
the well-known and widely used symmetric or circular gauge. If
$\xi=0$ or 1, we get the asymmetric or Landau gauge.

\unitlength=1in
\begin{figure}
\begin{center}
\begin{picture}(3,1.8)
\put(-0.7,-0.8){\psfig{file=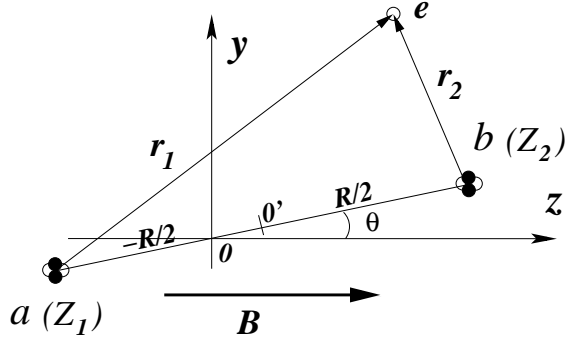,width=4.5in,angle=-90}}
\end{picture}
\end{center}
\caption{Geometrical setting for the system of two charged
         centers of the charges $Z_1$ and $Z_2$ placed on a line
         which forms the angle $\tha$ with magnetic line
         (inclined configuration), $0'$ is mid-point between
         $a$ and $b$.}
         \label{fig:1}
\end{figure}

To study the ground state of the Hamiltonian (\ref{ham-HeH}) for
the case of the $(He H)^{2+}$ we use the variational method with
the physically relevant trial function in a way similar to what
was done for the $H_2^+$ molecular ion in a strong magnetic field
(see \cite{Turbiner:2002}). A general recipe of the choice of the
trial function is presented in \cite{Tur}.  The trial function has
a form
\begin{equation}
\label{Psi-HeH}
 \Psi_{trial} = A_1 \psi_1 + A_2 \psi_2 \ ,
\end{equation}
where
\begin{subequations}
\label{psi123-HeH}
\begin{eqnarray}
 \psi_1 &=& {\large e^{-\al_1 Z r_1 - B  [\be_{1x} \xi x^2 +
 \be_{1y}(1-\xi) y^2]} +
 A e^{-\al_2 r_2 - B  [\be_{2x} \xi x^2 + \be_{2y}(1-\xi) y^2]}
} \ ,\\
 \psi_2 &=& {\large e^{-\al_3 Z r_1 -\al_4r_2
- B  [\be_{3x} \xi x^2 + \be_{3y}(1-\xi) y^2]}\ ,
}
\end{eqnarray}
\end{subequations}
here $\al_{1\ldots 4}$,
$\beta_{1x,1y},\beta_{2x,2y},\beta_{3x,3y}$, $A, A_{1,2}$ and
$\xi$ are variational parameters. The function $\psi_1$ simulates
the incoherent interaction of the electron with charged centers,
where the parameter $A$ "measures" an asymmetry in the interaction
of the electron with $\al$ and $p$. On the other side, $\psi_2$
describes the coherent interaction of the electron with $\al$ and
$p$. Considering the internuclear distance $R$ as a variational
parameter we have in total $10$ variational parameters (a free
normalization of the trial function (\ref{Psi-HeH}) allows us to
keep fixed one of the parameters $A_{1,2}$).

The result of calculations shows that the total energy surface
$E_T=E_T(B,R,\tha)$ for $B \gtrsim 1000$\,a.u. has global minimum at
$\tha=0^\circ$ and a finite internuclear distance $R=R_{eq}$ which
gives rise a valley when $\tha$ begins to vary. For smaller magnetic
fields there exists either no minimum or at most some irregularity.
Hence if the minima exists the optimal configuration for fixed
magnetic field $B \gtrsim 1000$\,a.u. always corresponds to zero
inclination, $\tha=0^\circ$ (parallel configuration), see for
illustration Fig.~\ref{fig:2-2}. Furthermore, for any fixed magnetic
field there exists a critical inclination $\theta_{cr}$ beyond of
which the minimum in the total energy curve at fixed inclination
disappears. It implies that the system $(He H)^{2+}$ does not exist
for inclinations larger than the critical inclination. For example,
for $B=10000$\,a.u. the critical angle $\tha_{cr} \approx 8^\circ$,
which is much smaller than $\tha_{cr}$ for $H_2^+$
\cite{Turbiner:2002}.

\unitlength=1in
\begin{figure}
\begin{center}
\begin{picture}(3.2,2.3)
\put(-0.4,-0.4){\psfig{file=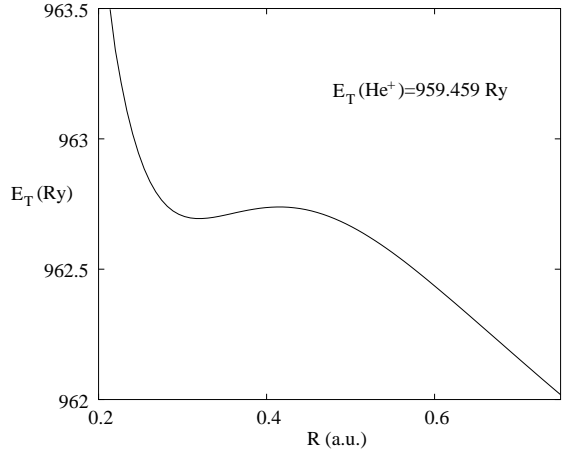,width=3.8in,angle=-90}}
\end{picture}
\end{center}
\caption{
  Total energy curve, $E_T$ viz. $R$, for the $(HeH)^{2+}$ molecular ion
  at $B=1000$\,a.u. in parallel configuration ($\tha=0^\circ$).}
  \label{fig:2-1}
\end{figure}

\unitlength=1in
\begin{figure}
\begin{center}
\begin{picture}(3.2,2.3)
\put(-0.4,-0.4){\psfig{file=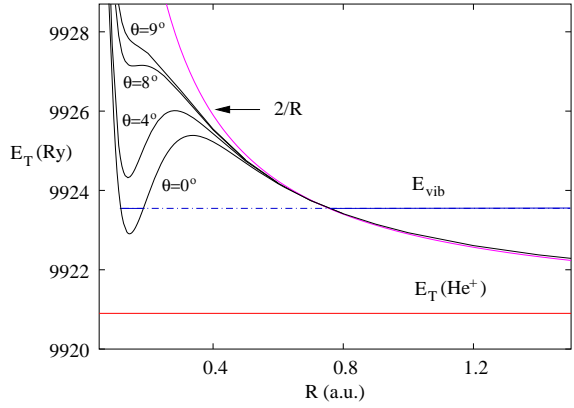,width=3.8in,angle=-90}}
\end{picture}
\end{center}
\caption{
  Total energy curve, $E_T$ viz. $R$, for the $(HeH)^{2+}$ molecular
  ion at $B=10000$\,a.u. for different inclinations; the dash line
  marks the total energy of the ${He}^+$ atomic ion, the solid
  horizontal line shows the lowest vibrational energy of $(HeH)^{2+}$.
  Dotted line corresponds to $2/R$.}
  \label{fig:2-2}
\end{figure}

\begin{figure}
\begin{center}
\begin{picture}(3.2,2.7)
\put(0,0.5){\psfig{file=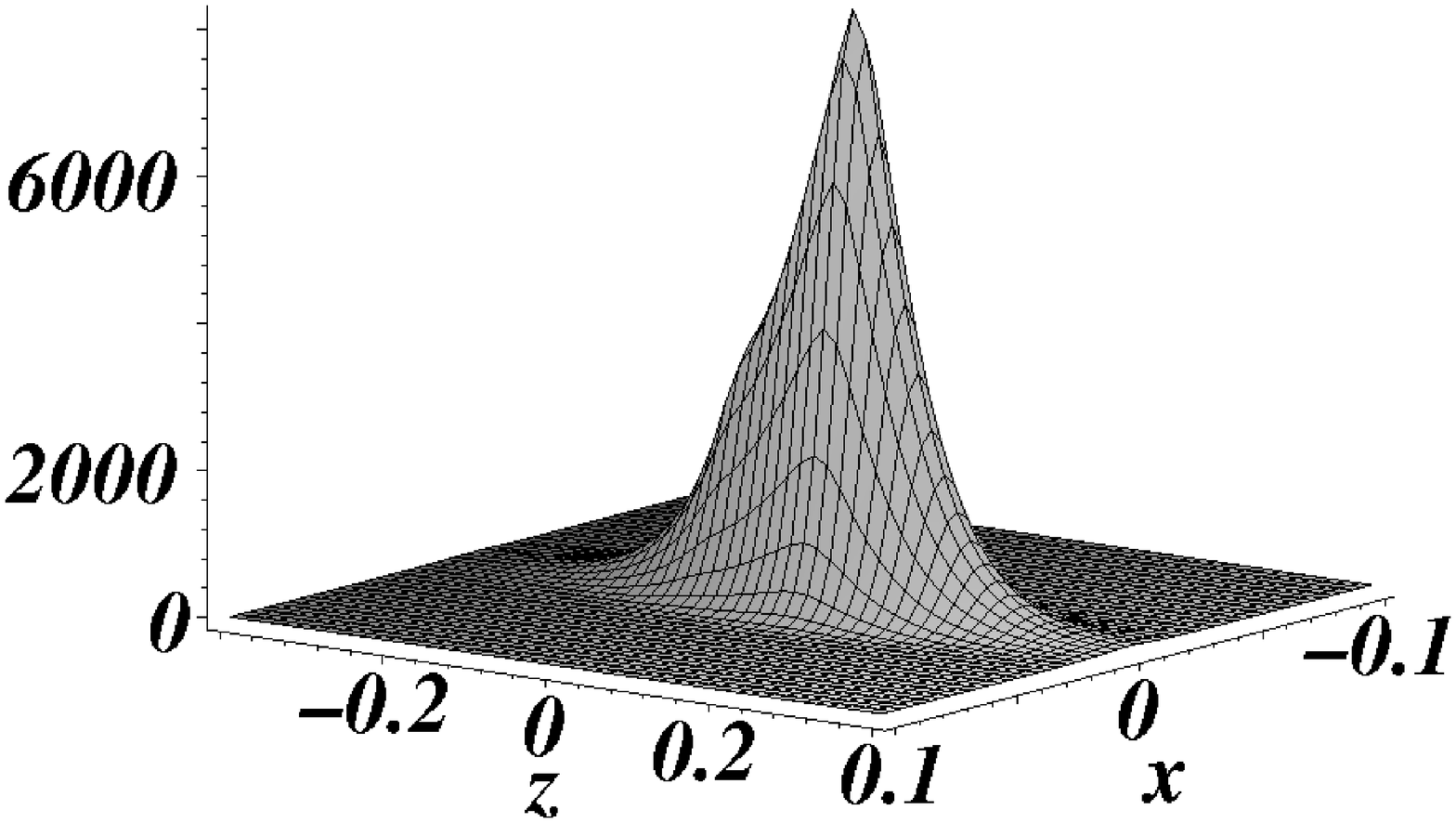,width=3in,angle=-90}}
\put(1.6,0.1){\large (a)}
\end{picture}
\begin{picture}(3.2,2.7)
\put(0,0.5){\psfig{file=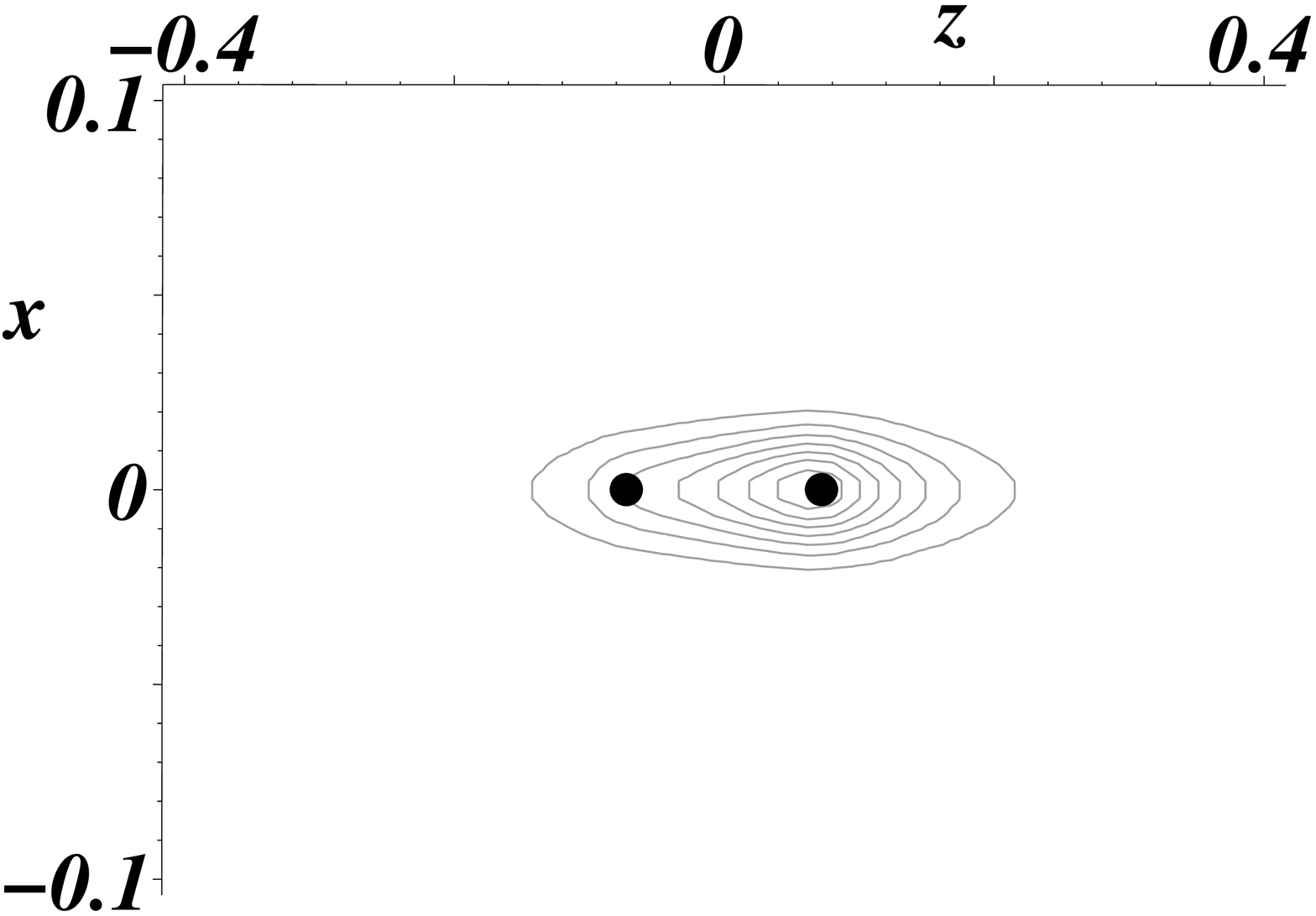,width=3in,angle=-90}}
\put(1.6,0.1){\large (b)}
\end{picture}
\end{center}
\caption{${(HeH)}^{2+}$ at $B=10000$\,a.u.: the electronic
 distribution $|\Psi(x,y=0,z)|^2/\int
 |\Psi(x,y,z)|^2 d^3 {\vec r}$ (a) and their contours (b)}
   \label{fig:2e}
\end{figure}

The total energy $E_T$, binding energy $E_b$ and equilibrium
distance $R_{eq}$ as well as the position and the height of the
barrier on the total energy curve of the system $(\al p e)$ in
parallel configuration for various magnetic fields are presented in
Table~\ref{Table:1}. Binding energy of $(He H)^{2+}$ is always
larger than the binding energy of the hydrogen atom. This implies
that ion $(He H)^{2+}$ does not decay to $H + \al$. In general, the
binding energy of $(He H)^{2+}$ grows very fast with the magnetic
field increase being smaller than the binding (ionization) energy of
the $He^{+}$ atomic ion
 \footnote[1]{The energies of the lowest energy states at the fixed
 magnetic quantum number $m$ for $H$ and ${He}^+, Li^{2+}$ atomic
 ions in a magnetic field $B$ are calculated using the 7-parameter
 trial function
 \[
    \psi_m = \rho^{|m|} e^{i m \phi} e^{\sqrt{\al_1^2 r^2 +
  (\ga_1 r^3 + \ga_2 \rho^2 r+\ga_3 \rho^3 + \ga_4 \rho r^2) B +
    \beta_1 B^2 \rho^4/16  + \beta_2 B^2 \rho^2 r^2/16}}
\]
  where $m$ is a magnetic quantum number and
  $\al,\ga_{1,2,3,4},\beta_{1,2}$ are variational parameters. For
  the hydrogen atom $H$ at $m=0$ (ground state) this trial function
  was successfully used in \cite{Potekhin:2001}.}.
It continues till $B \approx 3\times 10^{13}$\,G when these two
binding energies coincide. At larger magnetic fields $B \gtrsim
3\times 10^{13}$\,G the total energy of $(He H)^{2+}$ becomes lower
than the total energy of ${He}^+$ (see Table~I). It implies the
following picture. At the magnetic fields $B \lesssim 3\times
10^{13}$\,G the ion $(He H)^{2+}$ is unstable towards a decay to
$He^{+} + p$. However, at $B \gtrsim 3\times 10^{13}$\,G this decay
is forbidden and the exotic molecular ion $(He H)^{2+}$ becomes
stable \footnote{It is quite difficult to localize reliably a point
of transition in $B$ from a domain when decay is permitted to a
domain where it is forbidden (see Table I). An increase in accuracy
of the total energy of $(He H)^{2+}$ and/or $He^+$ can easily shift
this point.}.

In Figs.~\ref{fig:2-1}-\ref{fig:2-2} the total energy of $(He
H)^{2+}$ viz.  internuclear distance $R$ is shown at $B=1000$\,a.u.
and $B=10000$\,a.u., respectively, as an illustration.  At large
internuclear distances $R$, the behavior of the total energy is
defined by the interaction of $He^{+}$ and $p$.  It can be modeled
by the repulsive $2/R$-interaction term that is displayed by the
dotted line in Fig.~\ref{fig:2-2}, which is in good agreement with
the results of our variational calculations.

The equilibrium distance $R_{eq}$ shrinks down quite drastically,
reducing almost by a factor of three from $B=1000$\,a.u.  to
$4.414\times 10^{13}$\,G (see Table~\ref{Table:1}). Thus, the $(He
H)^{2+}$ ion becomes more compact as the magnetic field increases.
The position of the maximum $R_{max}$ of the total energy curve
$E_T(R)$ also reduces with magnetic field increase but not so
dramatically as for $R_{eq}$ (see Table~\ref{Table:1}).  Together
with the increase of the height of the barrier (see
Table~\ref{Table:1}), it indicates an increase in lifetime of $(He
H)^{2+}$ with magnetic field growth. The lifetime is always very
large. The electronic distribution is always a single-peak one with
the peak position shifted towards the charge $Z=2$.  As an
illustration the electronic distribution is presented for
$B=10000$\,a.u. in Fig.~\ref{fig:2e}. At $B \gtrsim 5 \times
10^{12}$\,G the lowest vibrational state of $(He H)^{2+}$ can also
exist (see Table~\ref{Table:1} and for illustration
Fig.~\ref{fig:2-2}). As for rotational states, we calculated the
energy of the lowest rotational state following the procedure
developed in \cite{Larsen:1982} for the $H_2^+$ molecular ion. The
rotational energy grows with a magnetic field increase being
systematically larger than vibrational ones unlike the field-free
case. For example, at $B=10000$\,a.u. the lowest rotational energy
is 6.72\,Ry while the lowest vibrational one is equal to 0.65\,Ry
only with the height of the barrier 2.481\,Ry (see Table~I). For
magnetic fields $B\gtrsim 5 \times 10^{12}$\,G the well keeps the
lowest vibrational state but not the lowest rotational state.

\begingroup
\squeezetable
\begin{table*}
  \label{Table:1}
  \caption{Ground state of the $(He H)^{2+}$ ion:  total energy $E_T$, binding
 energy $E_b=B-E_T$ and the equilibrium internuclear distance $R_{eq}$ \,(a.u.)
 in a magnetic field; the binding energy $E_b^{{He}^{+}}$ of the
 atomic ${He}^{+}$ ion is given ${}^{26}$. $R_{max}$ \,(a.u.) is a position
 of the maximum of the potential energy barrier, the height of the potential
 energy barrier $\De E^{max-min}$ and the lowest vibrational energy $E_{vib}^0$.}
  \begin{ruledtabular}
  \begin{tabular}{ccccccccl}
  \hline
 $B$  & $R_{eq}$ & $E_T$\,(Ry) & $E_b$\,(Ry) & $E_b^{{He}^{+}}$\,(Ry)
 & $R_{max}$ & $\Delta E^{max-min}$\,(Ry) & $E_{vib}^0$\,(Ry) & Trial Function \\
\hline
 1000\,a.u. & 0.316  & 962.635  & 37.365 & 40.40 & 0.424 & 0.061 & 0.13
 & Eq.(\ref{HeH-psi}) \\
            & 0.320  & 962.694  & 37.306 & --    & 0.415 & 0.045  & 0.12
 & Eq.(3) \cite{Turbiner:2005}\\
  2000\,a.u. & 0.239 & 1953.082 & 46.918 & --    & 0.405 & 0.428 & 0.25
 & Eq.(\ref{HeH-psi}) \\
             & 0.240 & 1953.172 & 46.828 & --    & --    & --    & --
 & Eq.(3) \cite{Turbiner:2005}\\
 $10^{13}$\,G & 0.185& 4195.693 & 59.626 & 61.99 & 0.375 & 1.197 & 0.41
 & Eq.(\ref{HeH-psi}) \\
             & 0.186 & 4195.830 & 59.489 & --    & 0.370 & 1.104 & 0.41
 & Eq.(3) \cite{Turbiner:2005}\\
  8000\,a.u. & 0.152 & 7927.677 & 72.323 & 73.84 & 0.350 & 2.216 & 0.59
 & Eq.(\ref{HeH-psi}) \\
             & 0.153 & 7927.765 & 72.135 & --    & --    & --    & --
 & Eq.(3) \cite{Turbiner:2005}\\
 10000\,a.u. & 0.142 & 9922.697 & 77.303 & 78.43 & 0.341 & 2.651 & 0.65
 & Eq.(\ref{HeH-psi}) \\
             & 0.143 & 9922.906 & 77.094 & --    & 0.338 & 2.481 & 0.65
 & Eq.(3) \cite{Turbiner:2005}\\
 14000\,a.u. & 0.129 & 13914.675& 85.325 & 85.74 & 0.330 & 3.43  & 0.77
 & Eq.(\ref{HeH-psi}) \\
             & 0.130 & 13914.919& 85.081 & --    & --    & --    & --
 & Eq.(3) \cite{Turbiner:2005} \\
$4.414\times 10^{13}$\,G
             & 0.119 & 18690.121& 92.858 & 92.53 & 0.318 &  4.22 & 0.88
 & Eq.(\ref{HeH-psi}) \\
             & 0.120 & 18690.398& 92.581 & --    & 0.315 &  3.98 & 0.87
 & Eq.(3) \cite{Turbiner:2005}\\
\hline
\end{tabular}
\end{ruledtabular}
\end{table*}
\endgroup

Once a conclusion about parallel configuration as optimal for the
ground state is drawn the existence of the states with different
magnetic quantum numbers (other than $m=0$) can be explored at
$\tha=0^\circ$. In order to do it, the following trial function was
chosen,
\begin{equation}
  \label{HeH-psi}
  \Psi_{trial} = \rho^{|m|}\,{\rm e}^{i m\phi}\Psi_0 =
  \rho^{|m|}\,{\rm e}^{i m\phi} \left( \psi_1 + \psi_2 \right) \,,
\end{equation}
where
\begin{subequations}
\begin{eqnarray}
  \label{eq:2}
  \psi_1 &=& A_1 {\rm e}^{-\al_1 Z r_1-\beta_1 \frac{B \rho^2}{4}}+
           A_2 {\rm e}^{-\al_2 r_2-\beta_2 \frac{B \rho^2}{4}}\,, \\
  \psi_2 &=& A_3 {\rm e}^{-\al_3 Z r_1-\al_4 r_2-\beta_3 \frac{B \rho^2}{4}}+
             A_4 {\rm e}^{-\al_5 Z r_1-\al_6 r_2-\beta_4 \frac{B \rho^2}{4}}\,,
\end{eqnarray}
\end{subequations}
(cf. (\ref{Psi-HeH})-(4)), here $\al_{1\ldots 6}$, $\beta_{1\ldots
4}$, $A_{1\ldots 4}$ are variational parameters, $Z=2$. Considering
the internuclear distance $R$ as a variational parameter we have in
total $14$ variational parameters (a free normalization of the trial
function (\ref{HeH-psi}) allows us to keep fixed one of the
parameters $A_{1\ldots 4}$). The calculations were carried out for
the magnetic quantum number $m=0,-1,-2$.

At $m=0$ (ground state) the function (\ref{HeH-psi}) leads to a
certain improvement in total and binding energies in comparison with
(3)-(4) (see Table~\ref{Table:1}). Other characteristics are changed
slightly except for the height of the barrier - it increases in
$\approx 10\%$ for all studied magnetic fields indicating the system
$(He H)^{2+}$ is more stable. This increase in the height of barrier
appears as a consequence of the fact that the value of the minimal
total energy decreases while the maximum of the barrier remained
almost unchanged in comparison with previous calculation
\cite{Turbiner:2005}. On Fig.~\ref{HeH-par} the variational
parameters of the trial function (\ref{HeH-psi}) are shown. It is
worth emphasizing the smoothness of their dependence on the magnetic
field strength.

\unitlength=1pt
\begin{figure}
 \caption{$(HeH)^{2+}$: variational parameters of the trial function
 (\ref{HeH-psi}) viz. $B$ for the $1\si$ state (ground state).}
  \label{HeH-par}
  \[
  \begin{array}{c}
  \begin{picture}(220,140)(0,5)
  \put(45,0){$4\cdot 10^{12}$}
  \put(115,0){$10^{13}$}
  \put(155,0){$2\cdot 10^{13}$}
  \put(200,0){$4\cdot 10^{13}$}
  \put(115,-14){$B(G)$}
  \put(12,140){$5$}
  \put(12,109){$0$}
  \put(5,75){$-5$}
  \put(0,42){$-10$}
  \put(0,12){$-15$}
  \put(130,136){$A_1$}
  \put(120,122){$A_3$}
  \put(110,100){$A_2$}
  \put(90,67){$A_4$}
  \put(10,155){\includegraphics*[width=2.2in,angle=-90]{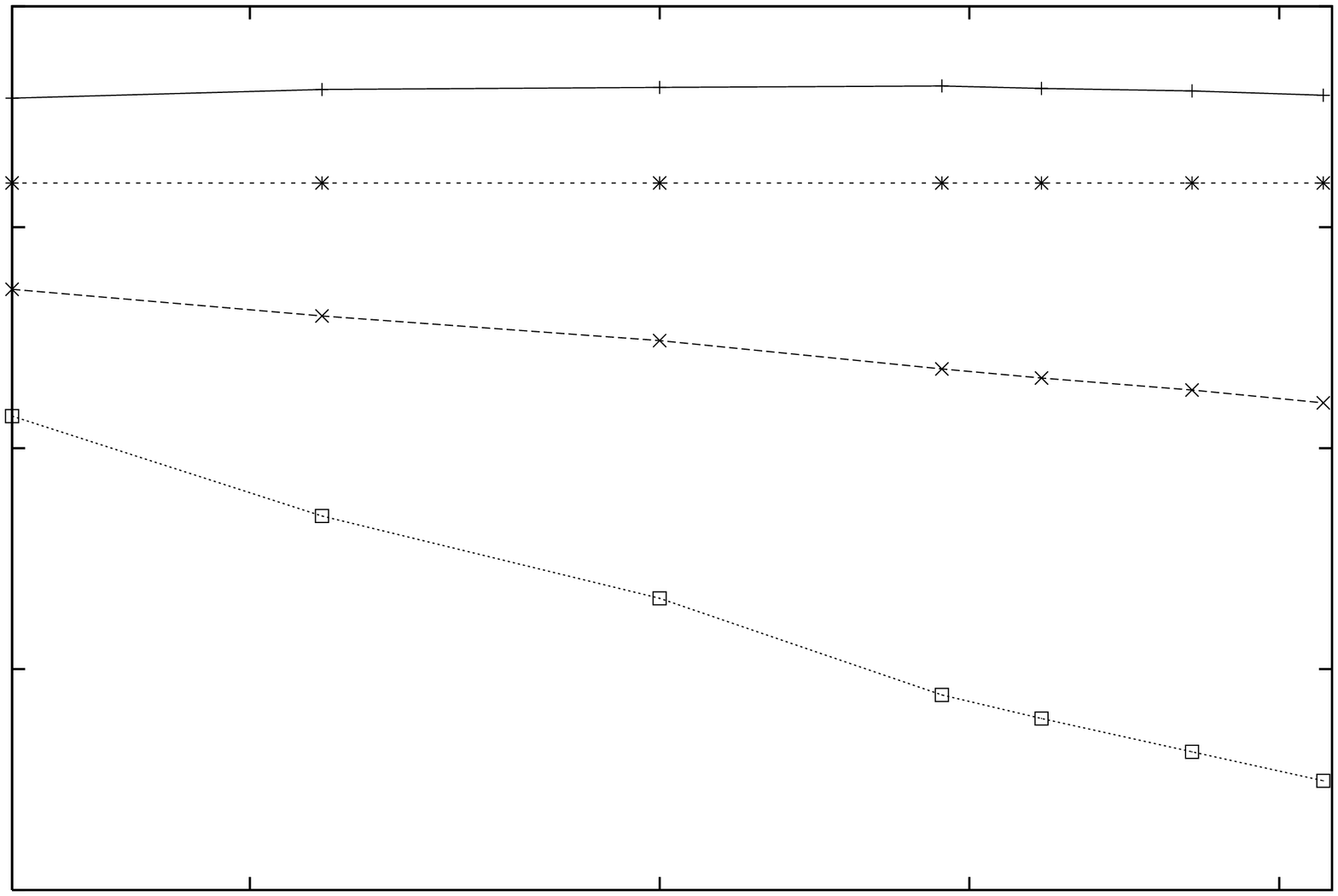}}
  \end{picture}
\\[30pt]
  \begin{picture}(220,140)(0,5)
  \put(45,0){$4\cdot 10^{12}$}
  \put(115,0){$10^{13}$}
  \put(155,0){$2\cdot 10^{13}$}
  \put(200,0){$4\cdot 10^{13}$}
  \put(115,-14){$B(G)$}
  \put(8,142){$35$}
  \put(8,108){$25$}
  \put(8,73){$15$}
  \put(14,42){$5$}
  \put(14,24){$0$}
  \put(5,9){$-5$}
  \put(-7,70){\rotatebox{90}{$[a.u.]^{-1}$}}
    \put(165,115){$\alpha_{4}$}
    \put(198,66){$\alpha_{2}$}
    \put(185,78){$\alpha_{1}$}
    \put(190,49){$\alpha_{5}$}
    \put(195,34){$\alpha_{6}$}
    \put(190,19){$\alpha_{3}$}
  \put(10,155){\includegraphics*[width=2.2in,angle=-90]{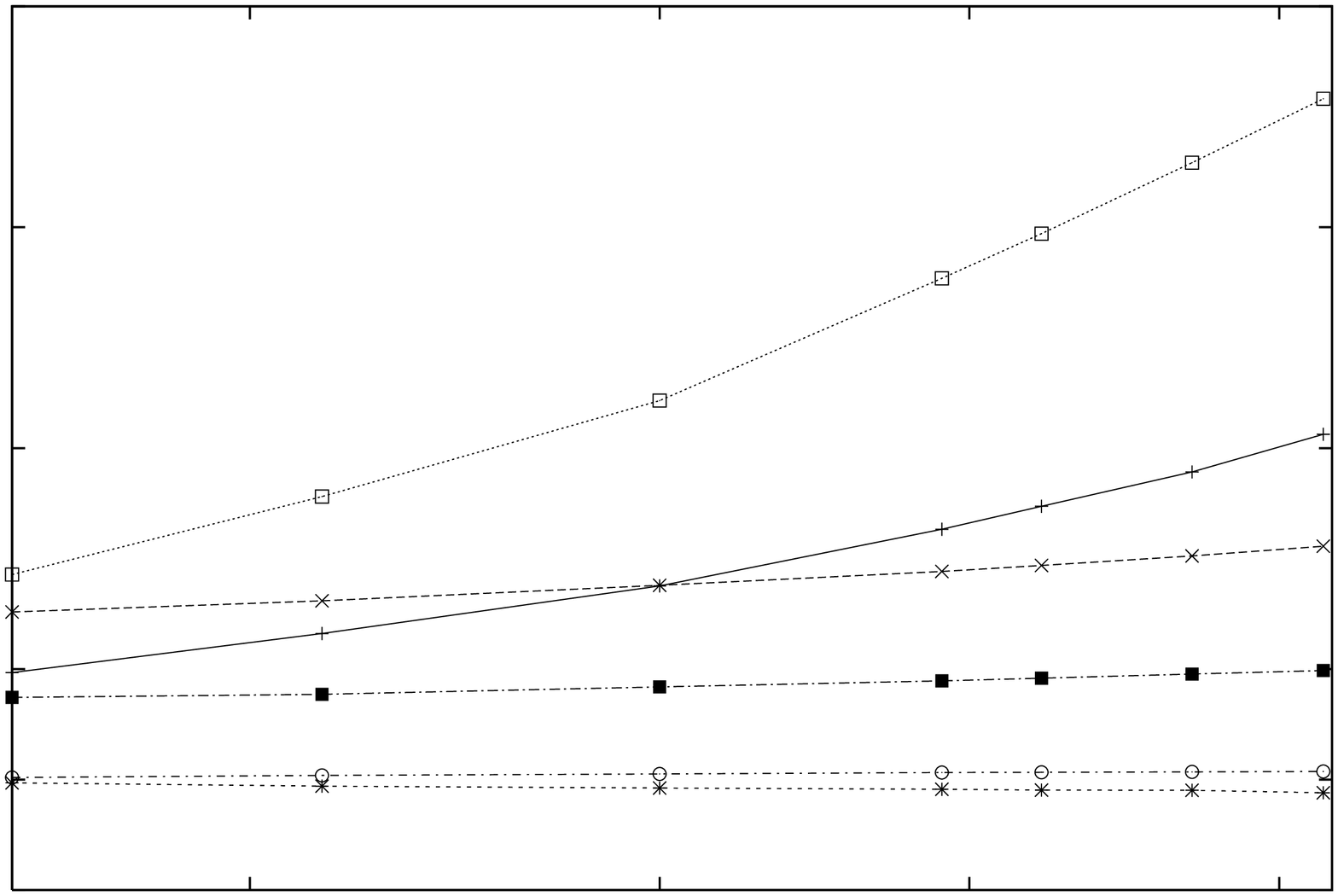}}
  \end{picture}
\\[30pt]
  \begin{picture}(220,140)(0,15)
  \put(45,12){$4\cdot 10^{12}$}
  \put(115,12){$10^{13}$}
  \put(155,12){$2\cdot 10^{13}$}
  \put(200,12){$4\cdot 10^{13}$}
  \put(115,-4){$B(G)$}
  \put(5,153){$1.2$}
   \put(5,110){$1.1$}
  \put(5,66){$1.0$}
  \put(5,22){$0.9$}
  \put(-7,80){\rotatebox{90}{$[a.u.]^{-1}$}}
    \put(95,115){$\beta_{1}$}
    \put(87,83){$\beta_{3}$}
    \put(127,68){$\beta_{2}$}
    \put(120,50){$\beta_{4}$}
  \put(10,167){{\includegraphics*[width=2.2in,angle=-90]{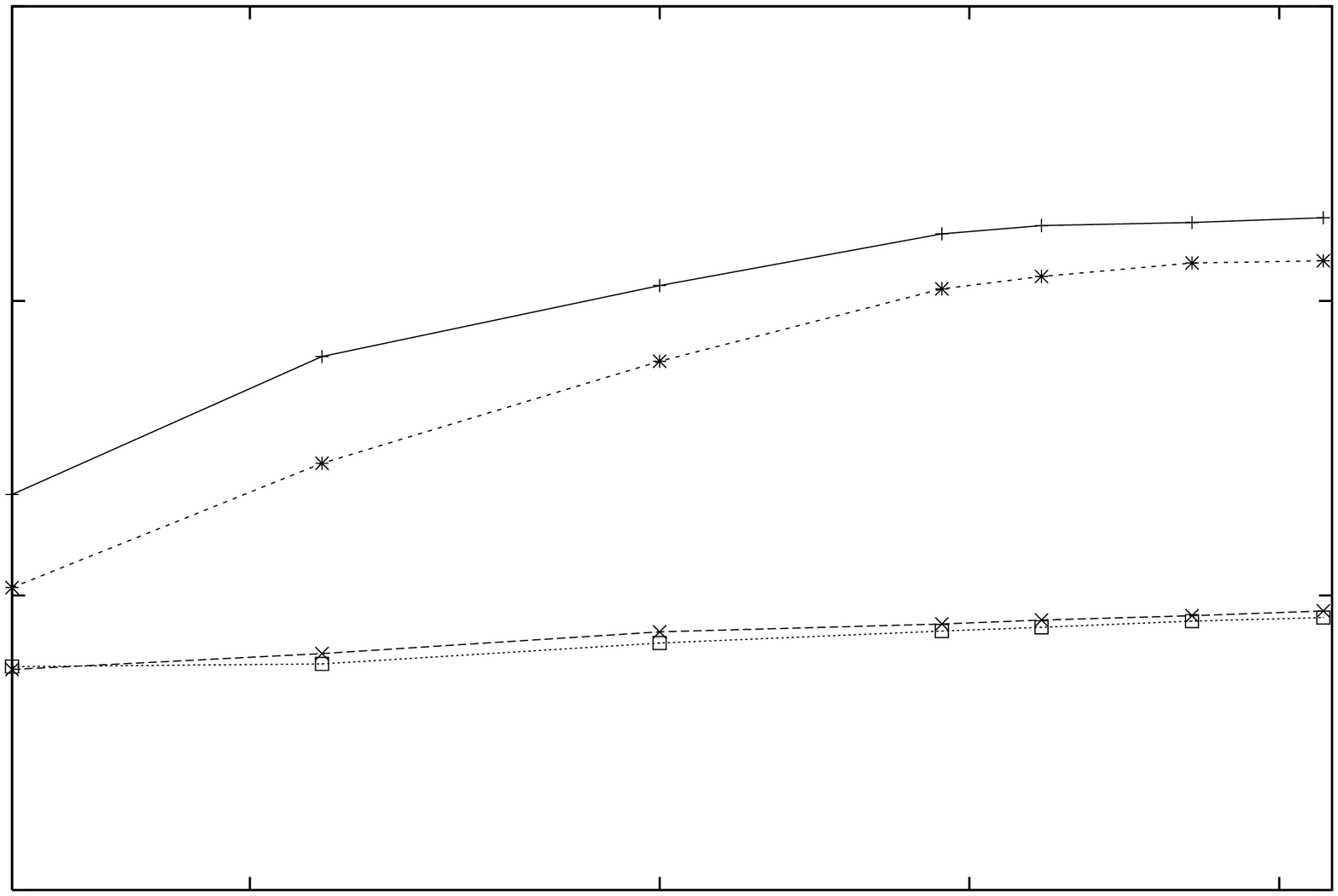}}}
  \end{picture}
\\[20pt]
  \end{array}
\]
\end{figure}

Using the trial function (\ref{HeH-psi}) it was found that the
nearest excited state of $(He H)^{2+}$ to the ground state is the
lowest in energy state at $m=-1$, the state $1\pi$ (see Table~II).
The state $1\pi$ of $(He H)^{2+}$ is stable towards a decay $(He
H)^{2+}(1\pi) \nrightarrow H (2p_{-1}) + \al$ but it can decay $(He
H)^{2+}(1\pi) \to He^+ (2p_{-1}) + p$. It is worth emphasizing that
the binding energies of $(He H)^{2+}(1\pi)$ and $He^+ (2p_{-1})$ are
very close. In Table~III the results of calculations of the $(He
H)^{2+}$ in $1\de$ state ($m=-2$) are shown. As in the states
$1\si,1\pi$ the ion $(He H)^{2+}$ in the state $1\de$ is stable
towards a decay to $H (3d_{-2}) + \al$ but it can decay to $He^+
(3d_{-2}) + p$. The difference in binding energies of $(He
H)^{2+}(1\de)$ and $He^+ (3d_{-2})$ is very small.

\begin{table*}
\begin{center}
\label{Table:1.1}
  \caption{
  The state $1\pi$ of the $(HeH)^{2+}$ in a strong magnetic field.
  $E_T$ is the total energy, $E_b=B-E_T$ is the binding energy,
  where $E_T^{He^+}$ is the total energy of the $He^{+}$ atomic
  ion in $2p_{-1}$ state,
  $R_{eq}$ \,(a.u.) the equilibrium internuclear distance.
  The binding energy $E_b^{He^+}$ of the $He^{+}$ atomic ion and the
  binding energy $E_b^{H}$ of the Hydrogen atom both in the $2p_{-1}$ state
  are shown ${}^{26}$. The binding energy of $E_b^{H_2^+}$ in $1\pi_u$ state
  (see \cite{Turbiner:2003}) is given for comparison.  }
  \begin{tabular}{ccccccc}
\hline
 $B$     &$R_{eq}$    & $E_T$\,(Ry)  & $E_b$\,(Ry) &
 $E_b^{\rm{He}^{+}} $\,(Ry) &$E_b^{\rm{H}} $\,(Ry) &$E_b^{\rm{H}_2^{+}}$\,(Ry)
   \\
\hline
$10^{13}$\,G & 0.263 & 4212.485  & 42.834 & 45.65 & 17.10 & 26.136 \\
 8000\,a.u. & 0.210  & 7947.374  & 52.626 & 55.13 & 20.32 & -  \\
10000\,a.u.  & 0.195 & 9943.497  & 56.503 & 58.83 & 21.56 & 34.068 \\
14000\,a.u.  & 0.175 & 13937.213 & 62.787 & 64.76 & 23.53 &  - \\
$4.414\times 10^{13}$\,G
             & 0.160 & 18714.251 & 68.727 & 70.31 & 25.36 & 41.090 \\
\hline
\end{tabular}
\end{center}
\end{table*}

\begin{table*}
\begin{center}
\label{Table:1.2}
  \caption{
  The state $1\de$ of the $(HeH)^{2+}$ in a strong magnetic field.
  $E_T$ is the total energy, $E_b=B-E_T$ is the binding energy,
  where $E_T^{He^+}$ is the total energy of the $He^{+}$ atomic ion in
  $3d_{-2}$ state, $R_{eq}$ \,(a.u.) the equilibrium internuclear distance.
  The binding energy $E_b^{He^+}$ of the $He^{+}$ atomic ion and the
  binding energy $E_b^{H}$ of the Hydrogen atom both in the $3d_{-2}$ state
  are shown ${}^{26}$. The binding energy of $E_b^{H_2^+}$ in $1\de_g$ state
  (see \cite{Turbiner:2003}) is given for comparison.  }
\begin{tabular}{ccccccc}
\hline
  $B$     & $R_{eq}$    & $E_T$\,(Ry)  & $E_b$\,(Ry) &
  $E_b^{\rm{He}^{+}} $\,(Ry)   & $E_b^{\rm{H}} $\,(Ry) &
  $E_b^{\rm{H}_2^{+}}$\,(Ry) \\
\hline
$10^{13}$\,G & 0.324 & 4219.204  & 36.115 & 38.93  & 14.77  & 22.19 \\
 8000\,a.u.  & 0.251 & 7955.361  & 44.639 & 47.34  & 17.70  & - \\
10000\,a.u.  & 0.231 & 9951.968  & 48.032 & 50.63  & 18.82  & 29.20 \\
14000\,a.u.  & 0.206 & 13946.450 & 53.550 & 55.94  & 20.60  & - \\
$4.414\times 10^{13}$\,G
             &0.187  & 18724.192 & 58.787 & 60.91  & 22.26  &  35.41 \\
\hline
\end{tabular}
\end{center}
\end{table*}

Finally, one can draw a conclusion that the exotic ion $(He H)^{2+}$
can exist in parallel configuration for magnetic fields $B\gtrsim 5
\times 10^{12}$\,G as a long-living quasi-stationary state with the
total energy which is slightly higher than the total energy of the
atomic ion $He^+$ \footnote{In domain $5 \times 10^{12} \lesssim
B\lesssim 3 \times 10^{13}$\,G where the first vibrational state
exists using Gamov theory (see e.g. \cite{BM}) we estimated the
lifetime of $(He H)^{2+}$}. However, $(He H)^{2+}$ becomes stable at
$B\gtrsim 3 \times 10^{13}$\,G and its two lowest energy electronic
states at $m=-1,-2$ can also exist. Their binding energies grow with
a magnetic field increase. However, the total energy difference of
the lowest corresponding states of $(He H)^{2+}$ and $He^+$ at the
same magnetic quantum number $m$ reduces as magnetic field grows.
For given magnetic field their binding energies are reduced with
magnetic quantum number decrease.

It is worth commenting that we attempted to study the existence of
the bound state of the system made from lithium nuclei, proton and
electron in the parallel configuration. The Hamiltonian is given by
(\ref{ham-HeH}) at $Z_1=3, Z_2=1$. We used the trial function
(3)-(\ref{psi123-HeH}) at $Z_1=3$. It turns out that at $B \gtrsim 7
\times 10^{14}$\,G a total energy curve displays a well-pronounced
minimum at zero inclination. Nevertheless, no definite conclusion
can be drawn, since these values of the magnetic field strength are
quite far beyond a domain of applicability of non-relativistic
consideration.

\section{The molecular ion $He_2^{3+}$}

Can two $\al$-particles be bound by one electron in a strong
magnetic field?\, Although, on the first sight the question sounds
surprising, it was just recently mentioned~\cite{Benguria:2004} that
based on semi-qualitative arguments the system $(\al \al e)$ can
develop a bound state for sufficiently strong magnetic fields.
Recently, we announced that this system can at $B \gtrsim
100\,\mbox{a.u.} (= 2.35\times 10^{11}\,\mbox{G})$
\cite{Turbiner:2005}. Our present goal is to make a detailed
analysis of the existence of the bound states of the system $(\al
\al e)$.

The Hamiltonian (\ref{ham-HeH}) at $Z_1=Z_2=Z=2$ describes two
infinitely heavy $\alpha-$particles situated along the line forming
the angle $\theta$ with the $z-$axis, and one electron placed in a
uniform constant magnetic field directed along the $z-$axis As a
method to explore the problem we use the variational procedure. The
ground state trial function is the same as a function which was
successfully used to explore the $H_2^+$ molecular ion in a strong
magnetic field for arbitrary inclination
\cite{Turbiner:2002,Turbiner:2003}. It has a form
\begin{equation}
\label{Psi-He2}
 \Psi_{trial} = A_1 \psi_1 + A_2 \psi_2 + A_3 \psi_3 \ ,
\end{equation}
with
\begin{subequations}
\label{psi123-He2}
\begin{eqnarray}
 \psi_1 &=& {\large e^{-\al_{1} Z (r_1 + r_2)}\,
e^{- B  [\be_{1x} \xi x^2 + \be_{1y}(1-\xi) y^2]}
} \\
 \psi_2 &=& {\large (e^{-\al_2 Z r_1} + e^{-\al_2 Z r_2})\,
e^{- B  [\be_{2x} \xi x^2 + \be_{2y}(1-\xi) y^2]}
}\\
 \psi_3 &=& {\large (e^{-\al_3 Z r_1 -\al_4 Z r_2} +
e^{-\al_4 Z r_1 -\al_3 Z r_2})\, e^{-
 B  [\be_{3x} \xi x^2 + \be_{3y}(1-\xi) y^2]}
}
\end{eqnarray}
\end{subequations}
(cf. (\ref{psi123-HeH})), where $\alpha_{1\ldots 4}$,
$\beta_{1x,1y},\beta_{2x,2y},\beta_{3x,3y}$, $A_{1\ldots 3}$ and
$\xi$ are variational parameters. It is worth mentioning that the
trial function (\ref{Psi-He2})-(\ref{psi123-He2}) has provided for
$H_2^+$ the accurate results for $B \gtrsim 10^{9}$\,G, which are
the most accurate at $B \gtrsim 10^{10}$\,G \cite{Turbiner:2003} so
far. Considering the internuclear distance $R$ as a variational
parameter we end up with fifteen variational parameters (the
normalization of the trial function (\ref{Psi-He2}) allows one to
keep fixed one of the $A_{1,2,3}$ parameters). Each function
$\psi_{1,2}$ is a modification of the celebrated Heitler-London or
Hund-Mulliken functions by multiplication by the corresponding
lowest Landau orbital (modified to the case of inclined
configuration), respectively. The functions $\psi_{1,2}$ describe
the coherent (incoherent) interaction of the electron with charged
centers, respectively. In turn, the function $\psi_3$ is a modified
Guillemin-Zener function. It can be also considered as a non-linear
superposition of the functions $\psi_{1,2}$.

\begingroup
\squeezetable
\begin{table*}
  \label{Table:4}
 \caption{Ground state of $He_2^{3+}$-ion: total energy $E_T$, binding
 energy $E_b$ and the equilibrium internuclear distance $R_{eq}$ \,(a.u.)
 in a magnetic field; the binding energy $E_b^{{He}^{+}}$ of the
 atomic ${He}^{+}$ ion is given ${}^{26}$. $R_{max}$ \,(a.u.) is a position of
 the maximum of the potential energy barrier, the height of the potential energy
 barrier $\De E^{max-min}$, the lowest vibrational $E_{vib}^0$ and rotational
 $E_{rot}^0$ energies in Ry.}
  \begin{ruledtabular}
  \begin{tabular}{cccccccccc}
\hline
 $B$     &$R_{eq}$    & $E_T$\,(Ry)  & $E_b$\,(Ry) &
 $ E_b^{\rm{He}^{+}} $\,(Ry)  & $E_b^{\rm{H}_2^{+}}$\,(Ry)
 & $R_{max}$ & $\Delta E^{max-min}$\,(Ry) & $E_{vib}^0$\,(Ry) & $E_{rot}^0$\,(Ry)\\
\hline
100\,a.u.    & 0.780          & 83.484   & 16.516    & 19.11 &
 10.291  & 1.02 &   0.033 &  0.026 & 0.297 \\
150\,a.u.    & 0.640          & 130.702  & 19.298    & --   &
 -  & - &   - &  - & - \\
200\,a.u.    & 0.565          & 178.455  & 21.545    & 24.11 &
 -  & - &   - &  - & - \\
300\,a.u.    & 0.480          & 274.866  & 25.134    & 27.56 &
 -  & - &   - &  - & - \\
$10^{12}$\,G & 0.420         & 396.864   & 28.668    & 30.87 &
 17.143 & 0.90 &  1.024  &  0.100 & 0.810  \\
1000\,a.u.   & 0.309         & 960.732   & 39.268    & 40.40 &
 22.779 & 0.82 &  2.466  &  0.169 & 1.41 \\
$10^{13}$\,G & 0.193         & 4190.182  & 65.137    & 61.99 &
 35.754 & 0.70 &  7.328  &  0.366 & 3.718  \\
10000\,a.u.  & 0.150      & 9913.767     & 86.233    & 78.43 &
 45.797 & 0.62 &  12.25  &  0.561 & 6.476 \\
$4.414\times 10^{13}$\,G & 0.126 & 18677.857 & 105.121 & 92.53 &
 54.502 & 0.58 &  17.19  &  0.739 & 9.751\\
\hline
\end{tabular}
\end{ruledtabular}
\end{table*}
\endgroup

We performed accurate variational calculations of the total energy
for the system $(\al \al e)$ in magnetic fields ranging $B\simeq
10^{10}\,G - 4.414\times 10^{13}$\,G for different inclinations. The
obtained results indicate that for a  magnetic field $B\gtrsim
100$\,a.u.  $(=2.35\times 10^{11}$\,G) the total energy surface
$E_T=E_T(\theta,R)$ displays clearly a global minimum for a
$\theta=0^\circ$ and a finite value of the internuclear distance
$R=R_{eq}$. It indicates to the existence of the exotic molecular
ion $He_2^{3+}$ in parallel configuration as optimal (see Table~IV).
The equilibrium distance shrinks drastically as a magnetic field
increases making the system more and more compact. The total and
binding (ionization) energies grow as a magnetic field increases. It
is interesting to make a comparison of the binding energy $E_b$ of
the $He_2^{3+}$ molecular ion with the binding energy of the
$He^{+}$ atomic ion $E_b^{{He}^{+}}$. For magnetic fields $B <
1000$\,a.u. the binding energy of the atomic ion ${He}^{+}$ is
larger than one of the molecular ion $He_2^{3+}$. Therefore,
$He_2^{3+}$ is unstable towards the decay $He_2^{3+} \rar He^{+} +
\al$. However, for $B\gtrsim 1000$\,a.u. the relation is inverted:
the binding energy of the atomic ion ${He}^{+}$ is smaller than the
molecular ion $He_2^{3+}$. The above-mentioned decay becomes
forbidden and the exotic ion $He_2^{3+}$ becomes stable (see
Table~IV). It is quite striking that the binding energy of the
$He_2^{3+}$ is about twice larger than the binding energy of the
$H_2^{+}$ molecular ion in the whole region of magnetic fields where
both ions coexist (see Table~IV).

A potential curve of the total energy of $He_2^{3+}$ in parallel
configuration ($\tha=0^\circ$) is always characterized by the
existence of a barrier (for illustration see
Fig.~\ref{fig:3-1}-\ref{fig:3-2}).  A position of the maximum of the
barrier $R_{max}$ is reduced with a magnetic field increase, but
reduction is not that sharp as for $R_{eq}$. The height of the
barrier $\De E^{max-min}$ grows very fast with increase of a
magnetic field, much faster than the binding energy (see Table~IV).
These two facts can be considered as an indication that the lifetime
of $He_2^{3+}$ grows quickly with magnetic field increase before
becoming infinite.

\unitlength=1in
\begin{figure}
\begin{center}
\begin{picture}(3.2,2.3)
\put(-0.4,-0.4){\psfig{file=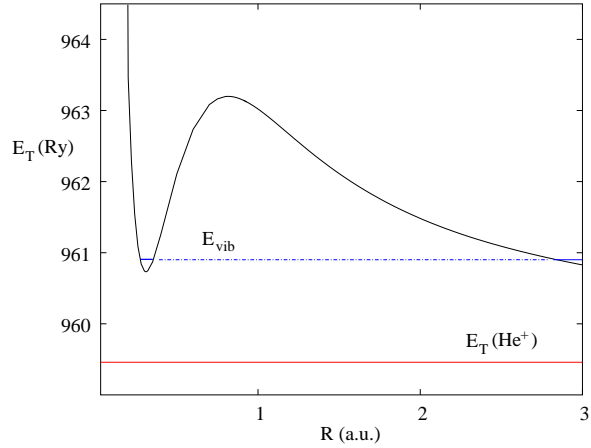,width=3.8in,angle=-90}}
\end{picture}
\end{center}
\caption{
  Total energy curve, $E_T$ viz. $R$, for the ${He_2}^{3+}$ molecular
  ion at $B=1000$\,a.u. in parallel configuration ($\theta=0^\circ$);
  the dotted-dash line marks the lowest vibrational energy, the solid line
  marks the total energy of the ${He}^+$ atomic ion.}
  \label{fig:3-1}
\end{figure}

\unitlength=1in
\begin{figure}
\begin{center}
\begin{picture}(3.2,2.3)
\put(-0.4,-0.4){\psfig{file=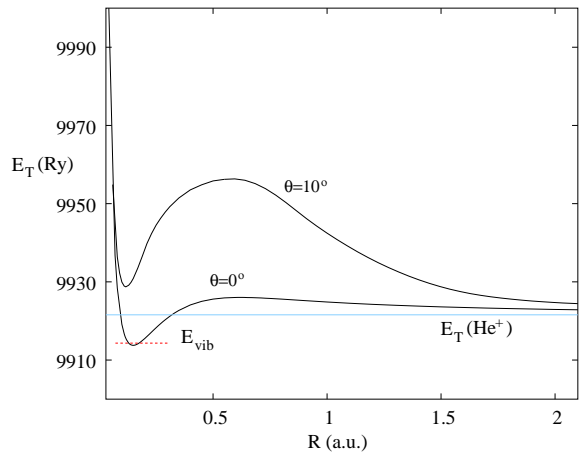,width=3.8in,angle=-90}}
\end{picture}
\end{center}
\caption{
  Total energy $E_T$ viz. the internuclear distance $R$ for the
  ${He}_2^{3+}$ at $B=10000$\,a.u. for parallel configuration,
  $\theta=0^\circ$ and for $\theta=10^\circ$ . Total energy of the
  atomic ion ${He}^+$ is shown by solid line, the dashed line marks
  the lowest vibrational energy.}
  \label{fig:3-2}
\end{figure}

We make a study of the lowest vibrational state and find that this
state exists for $B \gtrsim 100$\,a.u. (see Table~IV). The energy of
the lowest vibrational state (measured from the minimum of the
potential curve and written in Ry) is \footnote{There is a misprint
in the corresponding formula in \cite{Turbiner:2006}, p.388}
\[
E_{vib}^0 = \sqrt{\frac{\kappa}{\mu}}\ ,
\]
where $\mu=2\,m_{\rm proton}$ is the reduced mass of the system of
two $\al$-particles (considering $m_{\rm neutron} \approx m_{\rm
proton}=1836.15\,m_e$) and $\kappa$ is the curvature (in a.u.) of
the potential energy curve near the minimum. It seems evident that
the higher vibrational states can also exist for strong magnetic
fields. As for rotational states, we calculated the energy of the
lowest rotational state following the procedure developed in
\cite{Larsen:1982} for the $H_2^+$ molecular ion. The rotational
energy grows with a magnetic field increase being systematically
larger than vibrational ones unlike the field-free case. For
example, at $B=1000$\,a.u. the lowest rotational energy is 1.41\,Ry
while the lowest vibrational one is equal to 0.169\,Ry only with the
height of the barrier 2.466\,Ry; at $B=10000$\,a.u. the lowest
rotational energy is 6.02\,Ry while the lowest vibrational one is
equal to 0.561\,Ry only with the height of the barrier 12.25\,Ry
(see Table~IV). For magnetic fields $B\gtrsim 1000$\,a.u. the well
keeps both lowest rotational and vibrational states.

\unitlength=1in
\begin{figure}
\begin{center}
\begin{picture}(3.2,2.7)
\put(0,0.5){\psfig{file=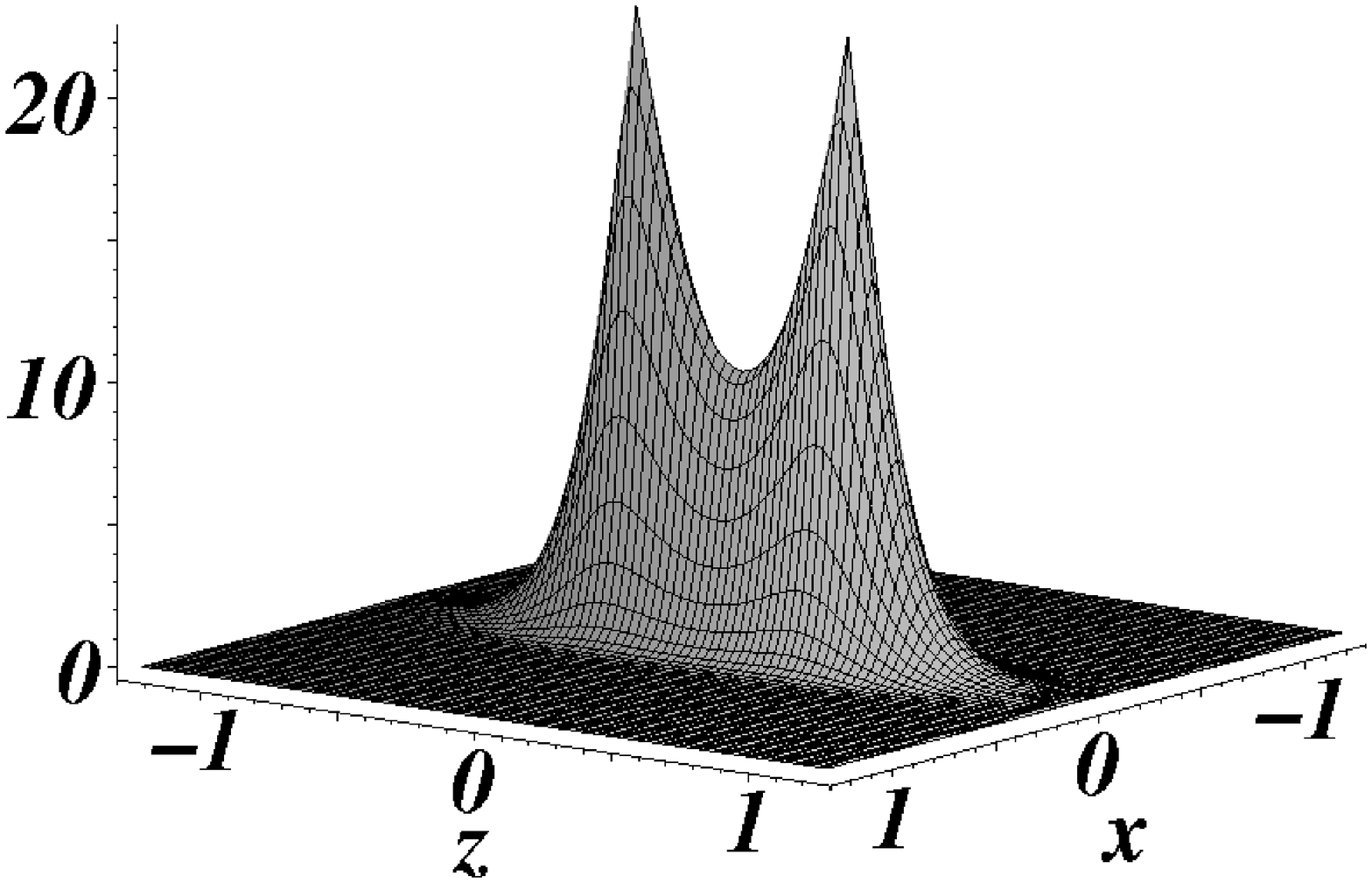,width=3in,angle=-90}}
\put(1.4,0.1){\large (a)}
\end{picture}
\begin{picture}(3.2,2.7)
\put(0,0.5){\psfig{file=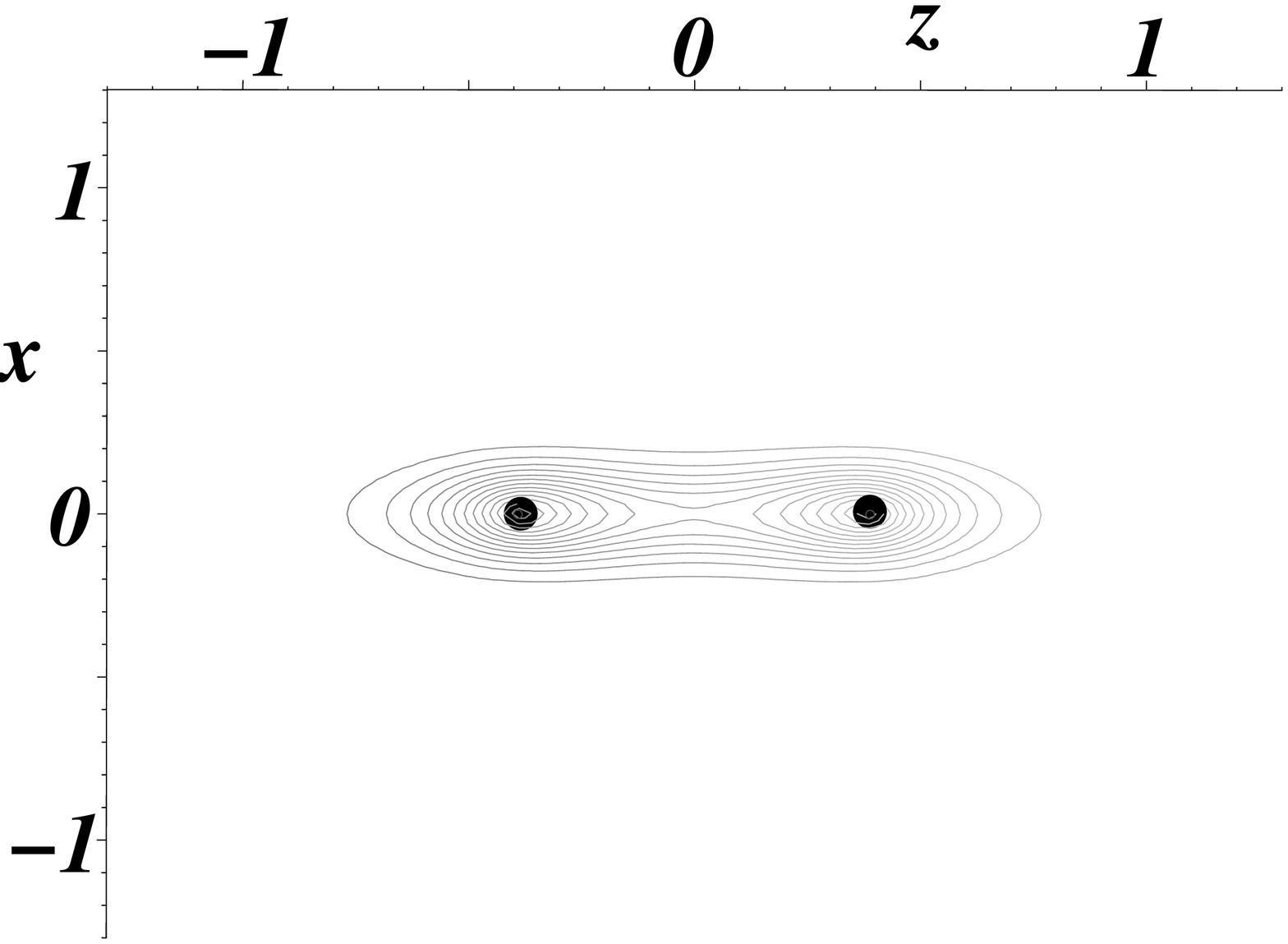,width=3in,angle=-90}}
\put(1.6,0.1){\large (b)}
\end{picture}
\end{center}
\caption{Electronic distribution $|\Psi(x,y=0,z)|^2/\int
    |\Psi(x,y,z)|^2 d^3 {\vec r}$ (a) and their contours (b)
    for the ${He}_2^{3+}$ molecular ion at $B=100$\,a.u.}
    \label{fig:4}
\end{figure}

\begin{figure}
\begin{center}
\begin{picture}(3.2,2.7)
\put(0,0.5){\psfig{file=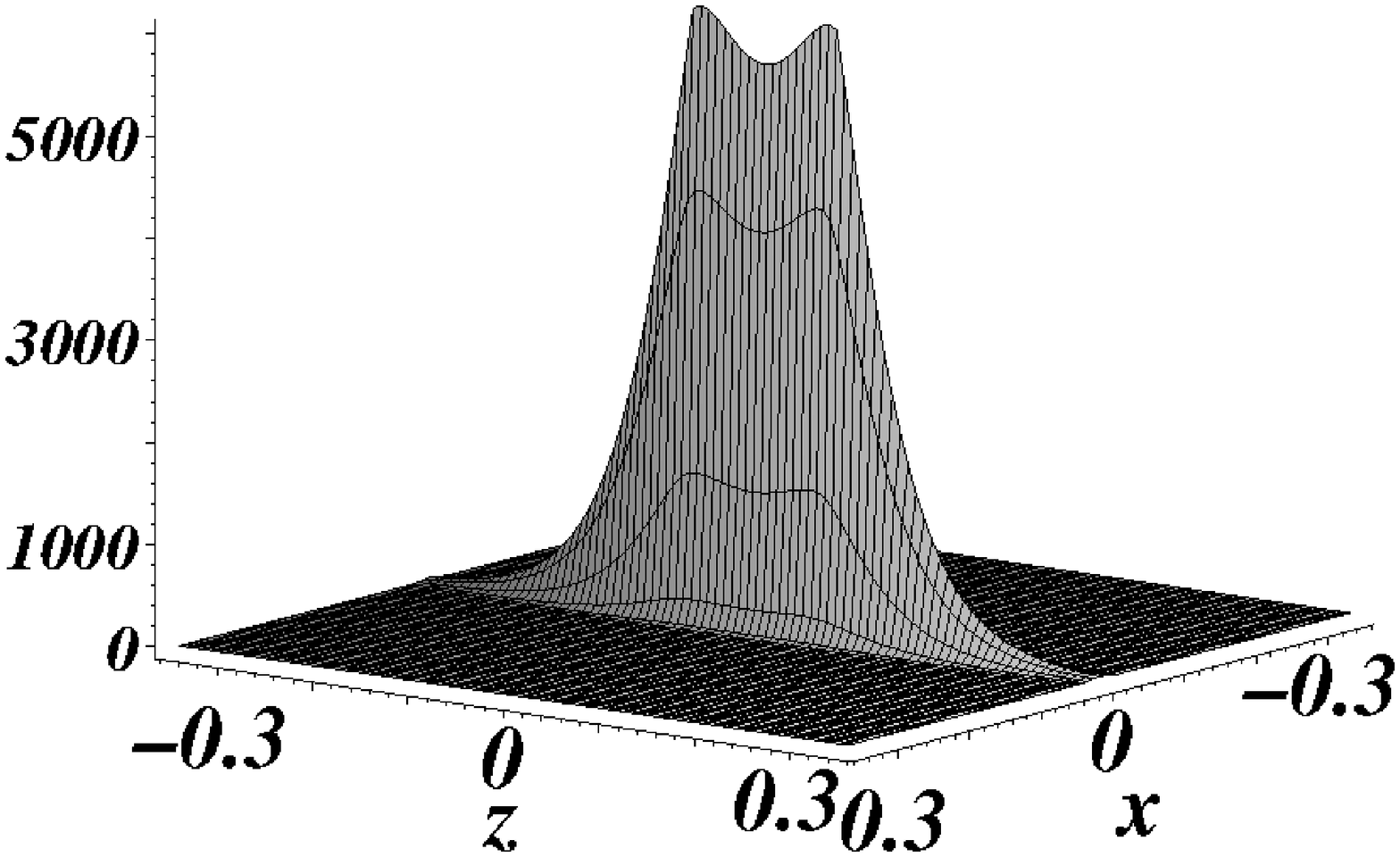,width=3in,angle=-90}}
\put(1.4,0.1){\large (a)}
\end{picture}
\begin{picture}(3.2,2.7)
\put(0,0.5){\psfig{file=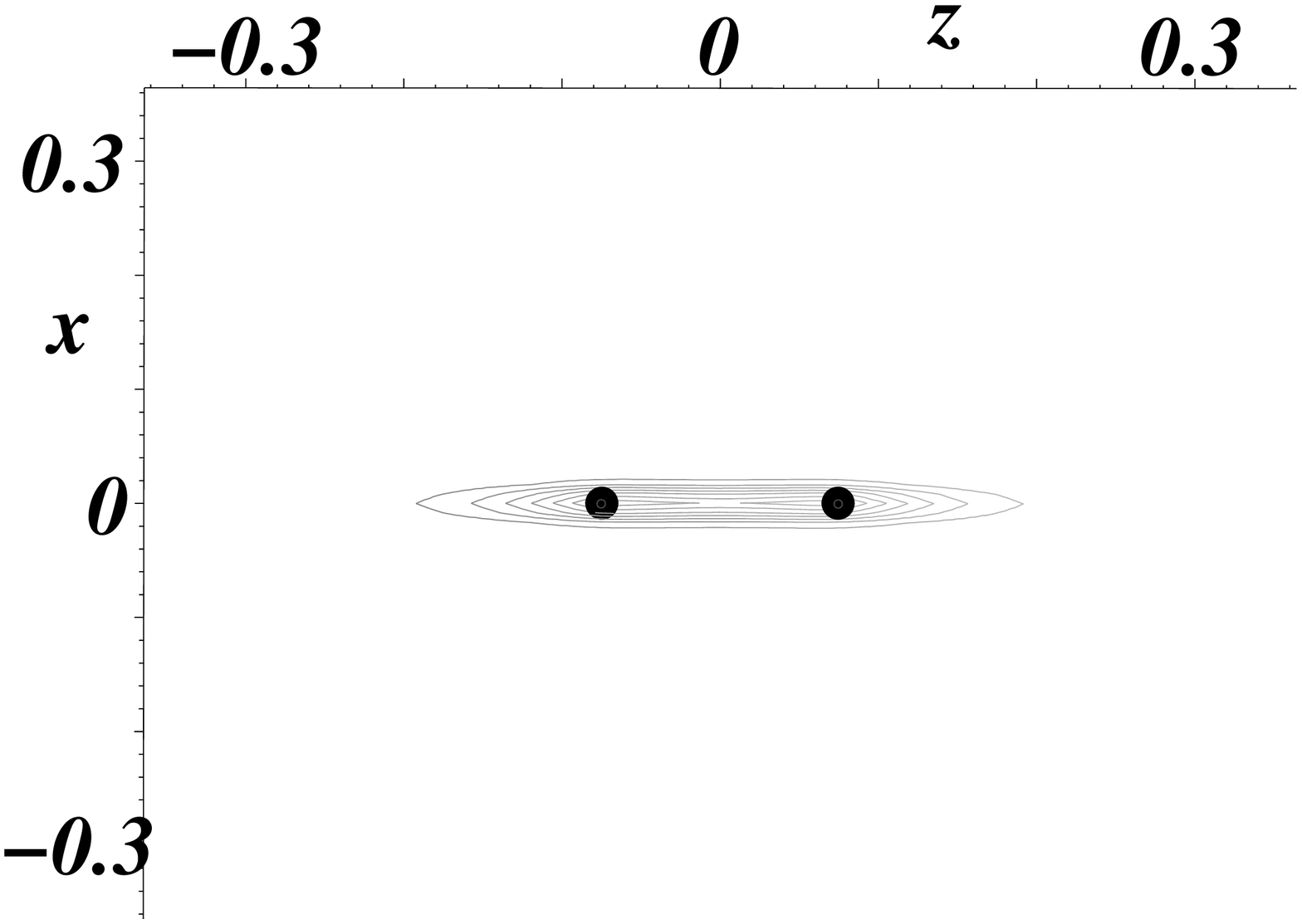,width=3in,angle=-90}}
\put(1.6,0.1){\large (b)}
\end{picture}
\end{center}
\caption{Electronic distribution $|\Psi(x,y=0,z)|^2/\int
   |\Psi(x,y,z)|^2 d^3 {\vec r}$ (a) and their contours
             (b) for the ${He}_2^{3+}$
                 molecular ion at $B=10000$\,a.u.}
   \label{fig:5}
\end{figure}

The electronic distributions for parallel configuration are illustrated in
Figs.~\ref{fig:4}-\ref{fig:5} for $B=100$\,a.u. and 1000\,a.u.,
respectively. In a domain of magnetic fields where the $He_2^{3+}$
ion can exist the electronic distribution has two clear pronounced
peaks near the positions of the charged centers. With a magnetic
field increase the peaks become less and less pronounced. Near the
Schwinger limit $B=4.414\times 10^{13}$\,G the electronic
distribution becomes almost uniform between the positions of
charges with exponential-type decay outside. It reminds a behavior
of the electronic distribution for the $H_2^+$ molecular ion in
parallel configuration in the domain $B=0 - 10^{11}$\,G. Average
transverse size of the distribution coincides approximately to the
Larmor radius.

The total energy surface $E_T=E_T(B,R,\theta)$ for $B \gtrsim
100$\,a.u. has global minimum at $\theta=0^\circ$ and some finite
$R=R_{eq}$. Hence, the optimal configuration for fixed magnetic
field $B \gtrsim 100$\,a.u. always corresponds to zero inclination,
$\theta=0^\circ$ (parallel configuration) in agreement with a
physics intuition (see for illustration Fig.~\ref{fig:3-2}). The
study shows that for magnetic fields $B\gtrsim 100$\,a.u. the total
energy at equilibrium grows with inclination, while the equilibrium
distance decreases with inclination. We find that for any fixed
magnetic field there exists a critical inclination beyond of which
the minimum in total energy curve at fixed inclination disappears.
It implies that the system $He_2^{3+}$ does not exist for
inclinations larger than the critical. This behavior is similar to
what was observed for the $H_2^+$ molecular ion
\cite{Turbiner:2002}. For example, for $B=10000$\,a.u. the critical
angle $\tha_{cr} \approx 28^\circ$, which is similar to $\tha_{cr}$
for $H_2^+$ \cite{Turbiner:2002}.

It can be concluded that the $He_2^{3+}$ ion can exist for $B
\gtrsim 100$\,a.u. in parallel configuration as optimal. This raises
a natural question about the existence of its excited states. Guided
by the results for $H_2^+$ where the lowest-lying excited states
were the lowest states but for magnetic quantum numbers $m=-1,-2$
(see \cite{Turbiner:2003}), we perform a study of $1\si_u,
1\pi_{u,g}, 1\de_{g,u}$ states. We use the trial functions similar
to those which were applied to study the excited states of $H_2^+$
\cite{Turbiner:2003}. They have a form of (7) multiplied by a factor
$\rho^{|m|} e^{i m \phi}$. It turns out that positive $z$-parity
states $1\pi_{u}, 1\de_{g}$ exist (see Tables~V-VI), while the
states of the negative $z$-parity $1\si_u, 1\pi_{g}, 1\de_{u}$ are
repulsive.

\begin{table}
\begin{center}
  \caption{$He_2^{3+}$: the results for the excited state $1\pi_u$.
   $E_T, E_b=B-E_T$ are the total and binding energies, $R_{eq}$ \,(a.u.)
   the equilibrium internuclear distance.  The binding energy
   $E_b^{\rm{He}^{+}}$ of the atomic ion $\rm{He}^{+}$ in the $2p_{-1}$
   state is given ${}^{26}$.}
  \label{Table:5.1}
  \begin{tabular}{cccccc} \hline
 $B$     &$R_{eq}$    & $E_T$\,(Ry)  & $E_b$\,(Ry) &
 $ E_b^{{He}^{+}} $\,(Ry)  & $E_b^{H_2^{+}}$\,(Ry)
  \\
\hline
200\,a.u.    & 0.936 & 185.971  & 14.029 & - & -    \\
300\,a.u.    & 0.743 & 283.474  & 16.526 & - & -    \\
$10^{12}$\,G & 0.631 & 406.525  & 19.007 & 21.483  & 11.902 \\
1000\,a.u.   & 0.448 & 973.421  & 26.579 & 28.749  & 16.126 \\
$10^{13}$\,G & 0.268 & 4209.722 & 45.597 & 45.645  & 26.136 \\
10000\,a.u.  & 0.203 & 9938.461 & 61.539 & 58.830  & 34.068 \\
$4.414\times 10^{13}$\,G
             & 0.168 & 18706.914 & 76.065 &70.312  & 41.090 \\
\hline
\end{tabular}
\end{center}
\end{table}

\begin{table}
\begin{center}
  \caption{$He_2^{3+}$: the results for the excited state $1\de_g$ of the
    molecular ion in a strong magnetic field.  $E_T, E_b=B-E_T$ are the
    total and binding energies, $R_{eq}$ \,(a.u.) the equilibrium
    internuclear distance. The binding energy $E_b^{{He}^{+}}$ of the atomic
    ${He}^{+}$ ion in the $3d_{-2}$ state is given ${}^{26}$. }
  \label{Table:5.2}
  \begin{tabular}{ccccccccc} \hline
 $B$     &$R_{eq}$   & $E_T$\,(Ry)  & $E_b$\,(Ry) &
 $ E_b^{He^{+}} $\,(Ry)  & $E_b^{H_2^{+}}$\,(Ry)
  \\
\hline
300\,a.u.    & 1.00  & 286.611  & 13.389  & -       & -    \\
$10^{12}$\,G & 0.798 & 410.089  & 15.443  & 17.853  &  9.861 \\
1000\,a.u.   & 0.545 & 978.234  & 21.766  & 24.127  & 13.488 \\
$10^{13}$\,G & 0.318 & 4217.409 & 37.910  & 38.928  & 22.194 \\
10000\,a.u.  & 0.238 & 9948.369 & 51.631  & 50.633  & 29.198 \\
$4.414\times 10^{13}$\,G
             & 0.195 & 18718.732 & 64.247 & 60.911  & 35.407 \\
\hline
\end{tabular}
\end{center}
\end{table}

It is worth demonstrating the plots of the variational parameters of
(7)-(8) viz the magnetic field. In general, the behavior is very
smooth but it is quite surprising that at $B \approx 1000$\,a.u.
where the $He_2^{3+}$-ion becomes stable the behavior of almost all
parameters is visibly changed.

\unitlength=1pt
\begin{figure}
 \caption{Variational parameters of the trial function
 (\ref{Psi-He2})-(8) as a function of the magnetic field strength
 $B$ for the $1\si_g$ state for the ${\rm He}_2^{3+}$ molecular ion.}
  \label{hehpar}
  \[
  \begin{array}{c}
  \begin{picture}(220,140)(0,5)
  \put(75,6){$10^{12}$}
  \put(155,6){$10^{13}$}
  \put(200,6){$4\cdot 10^{13}$}
  \put(115,-9){$B(G)$}
  \put(11,128){$1.5$}
  \put(11,90){$0.5$}
  \put(2,52){$-0.5$}
  \put(2,17){$-1.5$}
  \put(130,117){$A_3$}
  \put(150,45){$A_2$}
  \put(110,65){$A_1$}
  \put(10,155){\includegraphics*[width=2.2in,angle=-90]{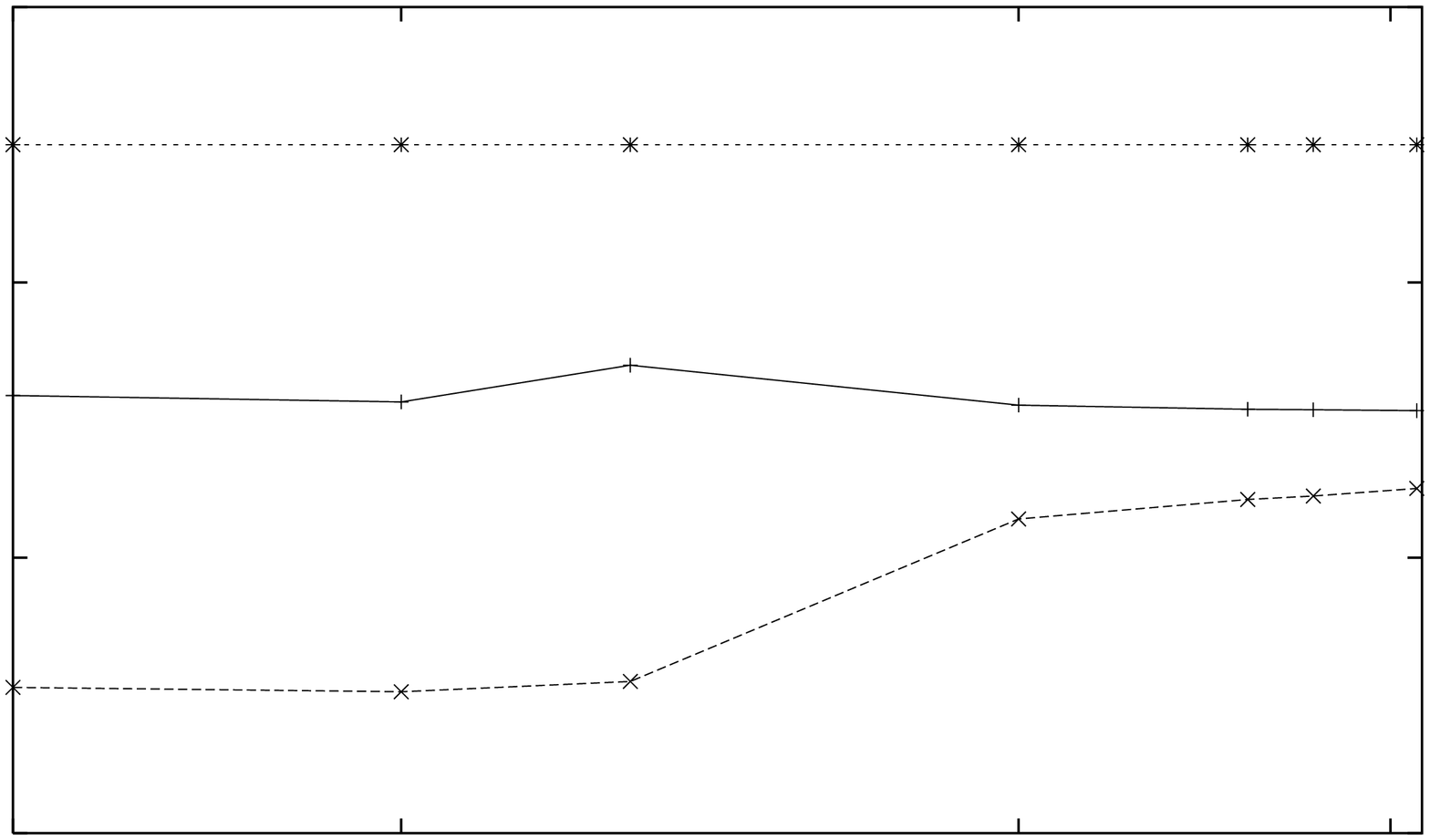}}
  \end{picture}
\\[30pt]
  \begin{picture}(220,140)(0,5)
  \put(75,5){$10^{12}$}
  \put(155,5){$10^{13}$}
  \put(200,5){$4\cdot 10^{13}$}
  \put(115,-10){$B(G)$}
  \put(14,113){$15$}
  \put(14,85){$10$}
  \put(20,57){$5$}
  \put(20,29){$0$}
  \put(-7,70){\rotatebox{90}{$[a.u.]^{-1}$}}
    \put(165,100){$\alpha_2$}
    \put(155,65){$\alpha_{3}$}
    \put(130,45){$\alpha_1$}
    \put(95,25){$\alpha_{4}$}
  \put(10,155){\includegraphics*[width=2.2in,angle=-90]{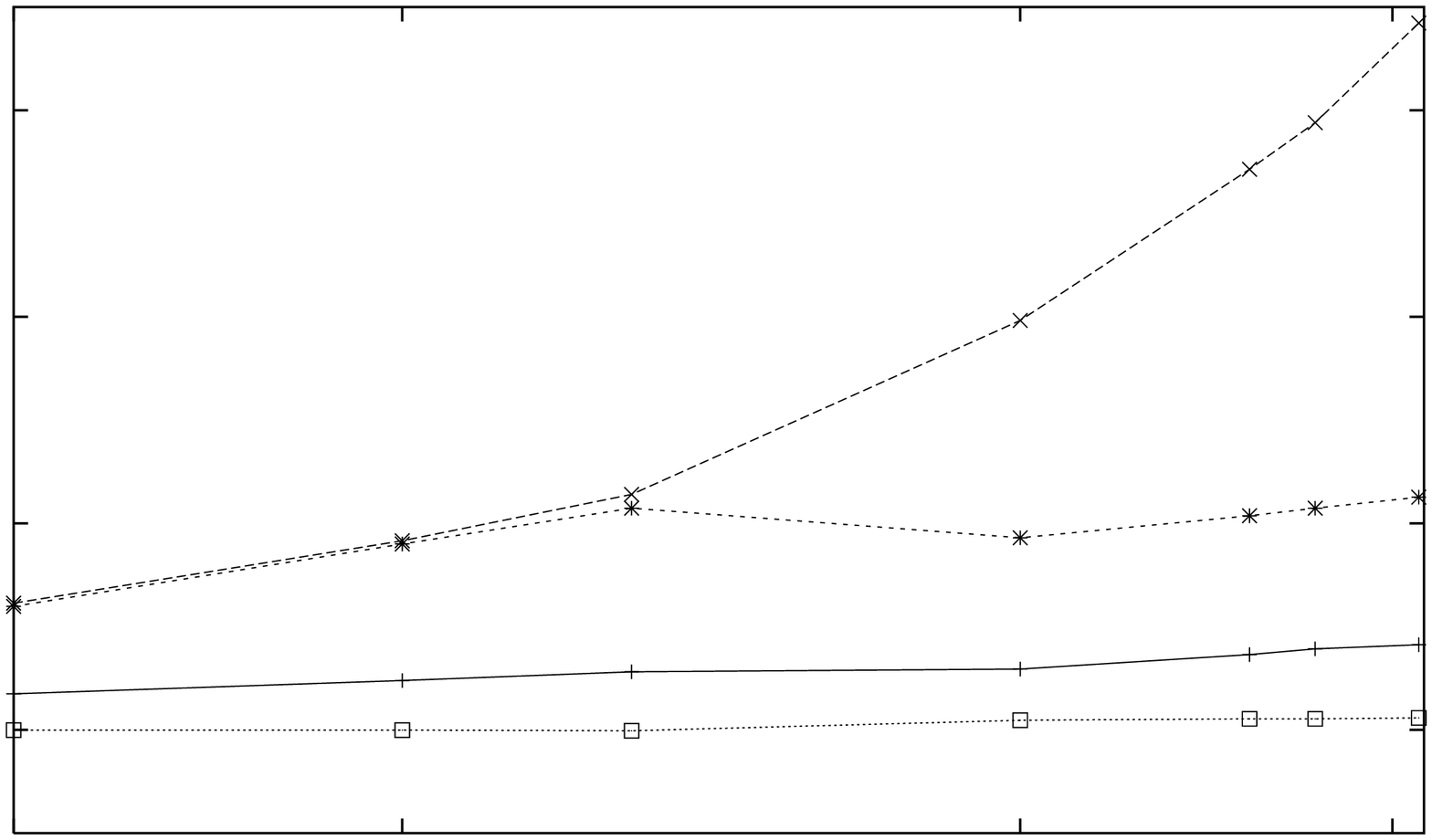}}
  \end{picture}
\\[30pt]
  \begin{picture}(220,140)(0,15)
  \put(75,18){$10^{12}$}
  \put(155,18){$10^{13}$}
  \put(200,18){$4\cdot 10^{13}$}
  \put(115,3){$B(G)$}
  \put(10,130){$1.6$}
  \put(10,85){$1.2$}
  \put(10,40){$0.8$}
  \put(-7,80){\rotatebox{90}{$[a.u.]^{-1}$}}
    \put(120,92){$\beta_{1}$}
    \put(170,88){$\beta_{2}$}
    \put(150,45){$\beta_{3}$}
  \put(10,167){{\includegraphics*[width=2.2in,angle=-90]{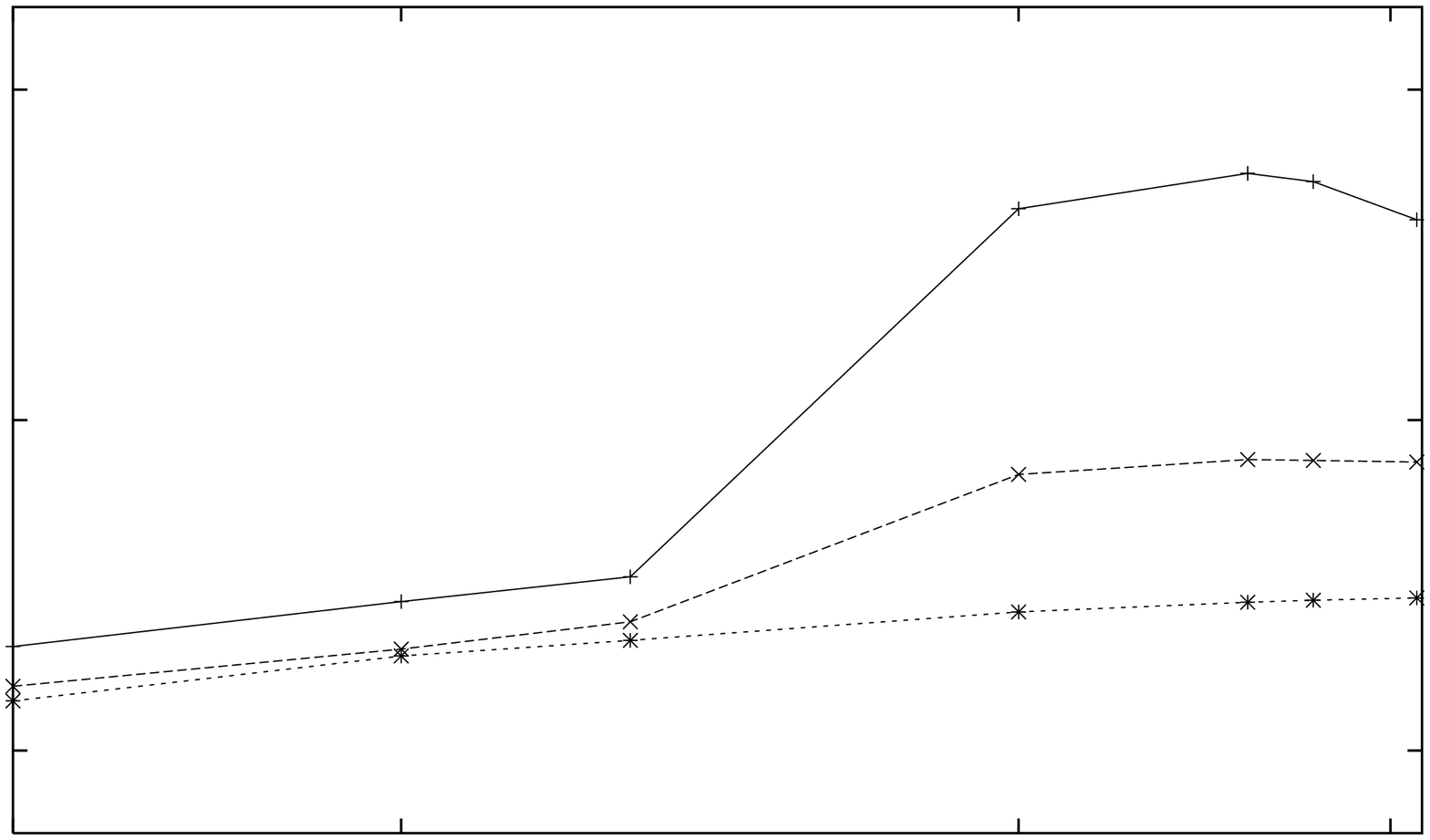}}}
  \end{picture}
\\[20pt]
  \end{array}
\]
\end{figure}

Finally, one can draw a conclusion that the exotic ion $He_2^{3+}$
can exist in parallel configuration for magnetic fields $B\gtrsim
100$\,a.u. At first, it exists as a long-living quasi-stationary
state with the total energy which is slightly higher than the total
energy of the atomic ion $He^+$ \footnote{In domain $100 \lesssim
B\lesssim 1000$\,a.u. where the first vibrational state exists using
Gamov theory (see e.g. \cite{BM}) we estimated the lifetime of
$He_2^{3+}$.}. Then $He_2^{3+}$ becomes stable at $B\gtrsim
1000$\,a.u. Making a comparison of the total energies of all
existing one-electron systems made out of protons and/or
$\al$-particles (see e.g. \cite{Turbiner:2006}) one can draw a
conclusion that at $B\gtrsim 1000$\,a.u. the exotic molecular ion
$He_2^{3+}$ has the lowest total energy! Two lowest energy
electronic states of $He_2^{3+}$ at $m=-1,-2$ can also exist. For
given magnetic field their binding energies are reduced with
magnetic quantum number decrease. To the contrary to $(He H)^{2+}$,
the total energy difference of the lowest corresponding states of
$He_2^{3+}$ and $He^+$ at the same $m$ increases as a magnetic field
grows.

\section{The systems ${Li}_2^{5+}$, $(H-He-H)^{3+}$, $(He-H-He)^{4+}$.}

\subsection{The molecular ion ${Li}_2^{5+}$}

Using the variational method as for the $He_2^{3+}$ molecular ion we
proceed to a question about the existence of the $Li_2^{5+}$
molecular ion in a magnetic field in parallel configuration. This
system contains two $Li$ nuclei and one electron. It is described by
the Hamiltonian (\ref{ham-HeH}) at $Z=3$ (see Fig.~\ref{fig:1} with
$Z_1=Z_2=3$). We used the same variational trial functions
(\ref{Psi-He2})-(\ref{psi123-He2}) as in the study of $He_2^{3+}$.
There was no indication to a minimum in the domain of applicability
of the non-relativistic approximation $B \leq 4.414\times
10^{13}$\,G. Hence, it is unlikely the ion $Li_2^{5+}$ can exist in
this domain. However, beyond of this domain, at $B \gtrsim 6 \times
10^{13}$\,G the total energy curve begins to display a minimum.
Although a contribution of relativistic corrections for this
magnetic field is unknown it seems natural to assume that they do
not change dramatically the situation. Keeping this in mind, we made
calculations for $B\ =\ 7 \times 10^{13}$\,G. The result is the
following. The total energy displays the well-pronounced minimum at
$R \sim 0.17$\,a.u. and the minimal energy is equal to
$29625.18$\,Ry. This minimum is stable towards small inclinations of
the molecular axis. The top of the potential barrier is located at
$R\sim 0.25$\,a.u. and the height of the barrier is $\De E \simeq
0.7$\,Ry. For the same magnetic field the total energy of the atomic
${Li}^{2+}$ ion is $E_T({Li}^{2+})\simeq 29597.14$\,Ry (see
${}^{25}$). Therefore, the system $(Li^{3+} Li^{3+} e)$ is highly
unstable towards dissociation ${Li}_2^{5+} \to {Li}^{2+} +
{Li}^{3+}$ with the large dissociation energy $\sim 28$\,Ry.

\subsection{The $(H-He-H)^{3+}$ and $(He-H-He)^{4+}$
 molecular ions}

\begin{figure}[tb]
\begin{center}
   \includegraphics*[width=2.2in,angle=-90]{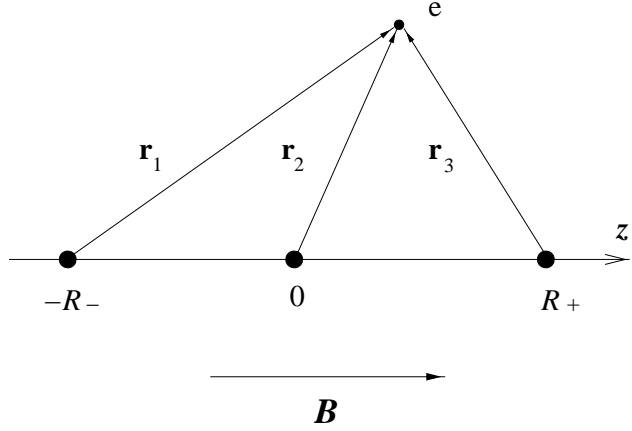}
    \caption{ Geometrical setting  for a problem of three charged centers
     situated along a magnetic line. The charged centers (marked by bullets)
     are situated on the $z$-line at distances $R_{\pm}$ from the central
     charge which is placed at the origin.}
      \label{fig:6}
\end{center}
\end{figure}

Let us check the existence of bound states of three-center linear
one-electron systems $(\al p p e)$ and $(\al \al p e)$. If they
exist, it would be an indication to the existence of the molecular
ions $(H-He-H)^{3+}$ and $(He-H-He)^{4+}$, correspondingly.
Following the experience of $H_3^{2+}$ (for a discussion see
\cite{Turbiner:2006}) it seems natural to assume that if above
systems would exist the optimal configuration should be linear and
parallel configuration, when all massive positively charged centers
are situated along a magnetic line. Since two centers are identical
there can appear two physical configurations depended on where the
third, non-identical particle is situated: either between two
identical particles (symmetric configuration) or as a side particle
(asymmetric configuration). Geometrical setting is presented in
Fig.~\ref{fig:6}. Below we state that following our calculations a
symmetric configuration when the non-identical particle is in
between the identical particles the bound state for each system can
exist, when the asymmetric configuration never leads to bound state.

The variational procedure is taken as a method to explore the
problem. The recipe of choice of trial functions is based on
physical relevance arguments and is described in full generality in
\cite{Tur}, to where the reader is addressed. The ground state trial
function is similar to one which was successfully used to explore
the $H_3^{2+}$ molecular ion in a strong magnetic field
\cite{Turbiner:1999,Turbiner:2004b} when the permutation symmetry
holds for two centers. The trial function for this case is of the
form
\begin{equation}
\label{Psi-HeH2}
 \Psi_{trial} = A_1 \psi_1 + A_2 \psi_2 + A_3 \psi_3 \ ,
\end{equation}
with
\begin{subequations}
\label{psi123-HeH2}
\begin{eqnarray}
 \psi_1 &=& {\large e^{-\al_{1}r_1 - \al_{2} r_2 - \al_{1}r_3}
e^{-
\frac{\beta_1}{4}B\rho^2}
} \\
 \psi_2 &=& {\large (e^{-\al_3 r_1} + e^{-\al_3 r_3}) e^{-\al_4 r_2}
e^{-
\frac{\beta_2}{4}B\rho^2}
}\\
 \psi_3 &=& {\large (e^{-\al_5 r_1 - \al_6 r_3} + e^{-\al_6 r_1 -\al_5 r_3})
e^{-\al_7 r_2
- \frac{\beta_3}{4}B\rho^2} }
\end{eqnarray}
\end{subequations}
where $\al_{1\ldots 7}$, $\beta_{1},\beta_{2},\beta_{3}$, and
$A_{1\ldots 3}$ are variational parameters. Taking the internuclear
distances $R_{\pm}$ as variational parameters we end up with $14$
variational parameters in total (the normalization of the trial
function (\ref{Psi-HeH2}) allows us to keep fixed one of the
$A_{1,2,3}$ parameters). The functions $\psi_{1}$ describes coherent
interaction of the electron with charged centers, when $\psi_{2}$
describes incoherent interaction of the electron with identical
centers.

\begingroup
\squeezetable
\begin{table*}
\centering
\squeezetable
 \caption{Ground states of different one-electron systems at
 $B=4.414\times 10^{13}$\,G. All energies are in (Ry) and $R_{\rm eq}$
 in (a.u.), see text. Total energies $E_T^{H}$ and $E_T^{{He}^+}$
 were calculated variationally, see ${}^{26}$.}
 \label{Table:7}
\begin{tabular}{|c|c|c|c|cccccc|}\hline
System
 & $E_T$ & $E_b$ & $R_{\rm eq}$ & $E_T^{H}$ &
 $E_T^{{H}_2^+}$ & $E_T^{{H}_3^{2+}}$ & $E_T^{{He}^+}$ &
$E_T^{(HeH)^{2+}}$ & $E_T^{He_2^{3+}}$
\\ \hline \hline
                         & & & & & & & & & \\
$(H-He-H)^{3+}$  & 18703.29 & 79.69 & 0.37 & & & & &  & \\
                        &          &       &      & 18750.48 &
                          18728.48 & 18727.75
& 18690.45 & 18690.12   & 18677.86 \\
$(He-H-He)^{4+}$ & 18712.22 & 70.76 & 0.34 & & & & & & \\
                         & & & & & & & & & \\
\hline
\end{tabular}
\end{table*}
\endgroup

Near the Schwinger limit $B=4.414\times 10^{13}$\,G for both systems
$(\al p p e)$ and $(\al \al p e)$ in symmetric configuration
$(p-\al-p)$ and $(\al-p-\al)$ the well-pronounced minimum in total
energy occurs at $R_+=R_-$ (see Table~\ref{Table:7}). In the same
time no indication to appearance of the minimum for asymmetric
configuration $(\al-p-p)$ and $(\al-\al-p)$ is found. For both
systems the discovered potential well is rather shallow, the
potential barrier heights are 0.34\,Ry and 0.26\,Ry, respectively.
But as a magnetic field increases it deepens quickly for both
systems. Both systems are quite compact (here the equilibrium
distance is defined as a distance between the end-situated heavy
particles $R_{\rm eq}= R_+ + R_-$, see Table~\ref{Table:7}). The
lifetimes of these systems are very short. The ion $(H-He-H)^{3+}$
is highly unstable towards decays $(HeH)^{2+} + p$, $(He)^{+} + 2p$,
but it does not decay to $H_2^{+}+\al$ or $H + \al + p$. In turn,
the ion $(He-H-He)^{4+}$ is highly unstable towards decays
$(HeH)^{2+} + \al$, $He_2^{3+} + p$, but it does not decay to
$He^{+} + \al + p$ or $H + 2\al$.

{\it Conclusion.} We presented theoretical arguments about a
possible existence of exotic ions $(HeH)^{2+}$ and ${He}_2^{3+}$ as
well as some indications to the existence of the exotic ions
${Li}_2^{5+}$, $(H-He-H)^{3+}$, $(He-H-He)^{4+}$. A striking
conclusion can be drawn that the exotic molecular ion $He_2^{3+}$ is
the most bound one-electron system among those made from protons
and/or $\al-$particles at strong magnetic field $B>1000$\, a.u. It
is then followed by two other $\al$-particle containing ions: the
atomic ion $He^{+}$ and the hybrid ion $(HeH)^{2+}$ - are
characterized by slightly larger total energies (see Tables I, IV,
VII). It is worth mentioning that $H$-atom - the only neutral
one-electron system is the least bound one. Both ions ${He}_2^{3+}$
and $(HeH)^{2+}$ can exist also in the excited states. In
Table~\ref{Table:7} a comparison of total energies of different
one-electron systems at the magnetic field $B=4.414\times
10^{13}$\,G is given. In very strong magnetic fields $B>3 \times
10^{13}$\,G the ions $H, H_2^+, H_3^{2+}, He^+, (HeH)^{2+},
{He}_2^{3+}$ are stable. In this domain of magnetic fields a
striking relation between the binding energies of ${He}_2^{3+}$ and
$H_3^{2+}$ exists: their ratio does not depend on magnetic field and
is equal to two. It may allow to explain the observed ratio of the
absorption features of the isolated neutron star 1E1207.4-5209
\cite{Sanwal:2002} by the helium-hydrogen content of its atmosphere
under a surface magnetic field $\approx 4.4 \times 10^{13}$\,G
\cite{Turbiner:2005m}.


\begin{acknowledgments}
  This work was supported in part by CONACyT grants
  {\bf 47899-E} (Mexico).
\end{acknowledgments}

\end{document}